\documentclass[aps,amsmath,notitlepage,twocolumn,amssymb,prb,longbibliography]{revtex4-1}

\usepackage[pdftex]{graphicx} 
\usepackage{epstopdf}
\usepackage{verbatim}
\usepackage{color}
\usepackage{subfigure}
\usepackage{tabularx}
\usepackage{amsfonts}
\usepackage{wasysym}
\usepackage{bm}

\usepackage[colorlinks=true,citecolor=red,urlcolor=blue]{hyperref}

\definecolor{darkred}{rgb}{0.847059, 0.141176, 0.164706}
\definecolor{darkgreen}{rgb}{0,0.4,0}
\definecolor{darkblue}{rgb}{0.254902, 0.411765, 0.882353}

\def\CH{{\cal H}}

\newcommand{\nn}{\nonumber\\}

\newcommand{\bra}[1]{\ensuremath{\langle#1|}}
\newcommand{\ket}[1]{\ensuremath{|#1\rangle}}

\newcommand{\rz}{\right]}
\newcommand{\lz}{\left[}

\newcommand{\dg}{\dagger}
\newcommand{\pg}{{\phantom{\dagger}}}

\newcommand{\ua}{\uparrow}
\newcommand{\da}{\downarrow}
\newcommand{\p}{\prime}

\renewcommand{\arraystretch}{1.5}

\newcolumntype{C}[1]{>{\centering\let\newline\\\arraybackslash\hspace{0pt}}m{#1}}

\begin{document}
\title{The spinon Fermi surface U(1) spin liquid in a spin-orbit-coupled \\triangular lattice Mott insulator YbMgGaO$_4$}
\author{Yao-Dong Li$^{1}$}
\author{Yuan-Ming Lu$^2$}
\author{Gang Chen$^{1,3}$}
\email{gangchen.physics@gmail.com}
\affiliation{$^{1}$State Key Laboratory of Surface Physics, Department of Physics,
Center for Field Theory \& Particle Physics,
Fudan University, Shanghai, 200433, China}
\affiliation{$^{2}$Department of Physics, The Ohio State University,
Columbus, OH, 43210, United States}
\affiliation{$^{3}$Collaborative Innovation Center of Advanced Microstructures,
Nanjing, 210093, China}

\date{\today}

\begin{abstract}
Motivated by the recent progress on the spin-orbit-coupled triangular 
lattice spin liquid candidate YbMgGaO$_4$, we carry out a systematic 
projective symmetry group analysis and mean-field study of candidate 
U(1) spin liquid ground states. Due to the spin-orbital entanglement 
of the Yb moments, the space group symmetry operation transforms both 
the position and the orientation of the local moments, and hence brings 
different features for the projective realization of the lattice 
symmetries from the cases with spin-only moments. Among the eight 
U(1) spin liquids that we find with the fermionic parton construction, 
only one spin liquid state, that was proposed and analyzed 
in Yao Shen,~\emph{et.~al.}, Nature 540, 559-562 (2016) and labeled as 
U1A00 in the present work, stands out and gives a large spinon Fermi 
surface and provides a consistent explanation for the spectroscopic 
results in YbMgGaO$_4$. Further connection of this spinon Fermi 
surface U(1) spin liquid with YbMgGaO$_4$ and the future directions 
are discussed. Finally, our results may apply to other spin-orbit-coupled 
triangular lattice spin liquid candidates, and more broadly, 
our general approach can be well extended to spin-orbit-coupled 
spin liquid candidate materials. 
\end{abstract}

\maketitle

\section{Introduction}
\label{sec1}

The interplay between strong spin-orbit coupling (SOC) 
and strong electron correlation has attracted a significant attention 
in recent years~\cite{WCKB}. At the mean time, the abundance of strongly 
correlated materials with $5d$ and $4f$ electrons, such as iridates and 
rare-earth materials~\cite{WCKB,rau2016spin},
brings a fertile arena to explore various emergent and exotic phases
that arise from such an interplay~\cite{Jackeli2009,Chen2008,PhysRevLett.105.027204,pesin2010mott,Onoda2010,PhysRevLett.108.037202,Ross2011,Huang2014,Chen2011,TbTiO,Chen2013,YbTiO2012,YbTiO2,PhysRevB.86.235129,PhysRevB.94.075146,PhysRevB.89.014424,PhysRevB.86.104412,PhysRevB.89.045117,PhysRevB.83.094411,YbTiO2012b,QSI1,li2016kitaev,PhysRevB.94.205107,0034-4885-77-5-056501,li2016octupolar,NdZrO1,PhysRevB.86.075154,PhysRevB.88.144402,Chen2010,Curnoe2008}. 
The recently discovered quantum spin
liquid (QSL) candidate YbMgGaO$_4$~\cite{srep}, where the rare-earth Yb
atoms form a perfect triangular lattice, is an ideal system
that involves strong spin-orbital entanglement in the {\it strong Mott insulating
regime} of the Yb electrons~\cite{YueshengPRL,YaodongPRB,YaoShenNature,
Martin2016,YueshengmuSR,Yaodong201608,PhysRevB.94.201114,Yaodong201703}.

In YbMgGaO$_4$, the thirteen $4f$ electrons of the Yb$^{3+}$ ions
are well localized and form a spin-orbit-entangled total moment
${\boldsymbol J}$ with ${J=7/2}$~\cite{YueshengPRL,YaodongPRB}.
The eight-fold degeneracy of the ${J=7/2}$ moment is further split by the
$D_{3d}$ crystal electric fields. The resulting ground state Kramers
doublet of the Yb$^{3+}$ ion, whose two-fold degeneracy is protected by 
the time-reversal symmetry, is well separated from the excited doublets and is
responsible for the low-temperature magnetic properties of YbMgGaO$_4$.
No signature of time-reversal symmetry breaking is observed for YbMgGaO$_4$
down to the lowest measured temperature~\cite{YueshengmuSR,Martin2016,YaoShenNature}.
Applying the recent theoretical result on spin-orbit-coupled 
Mott insulators~\cite{PNAS}, two of us and collaborators have 
proposed YbMgGaO$_4$ to be the first QSL candidate in the spin-orbit-coupled 
Mott insulator with odd electron fillings~\cite{YueshengPRL,YaodongPRB,YaoShenNature,Yaodong201608}.
More broadly, YbMgGaO$_4$ represents a new class of rare-earth materials 
where the strong spin-orbit entanglement of the local moments
meets with the geometrical frustration of the triangular lattice
such that exotic quantum phases may be stabilized.

\begin{figure}[b]
\centering
\includegraphics[width=7.2cm]{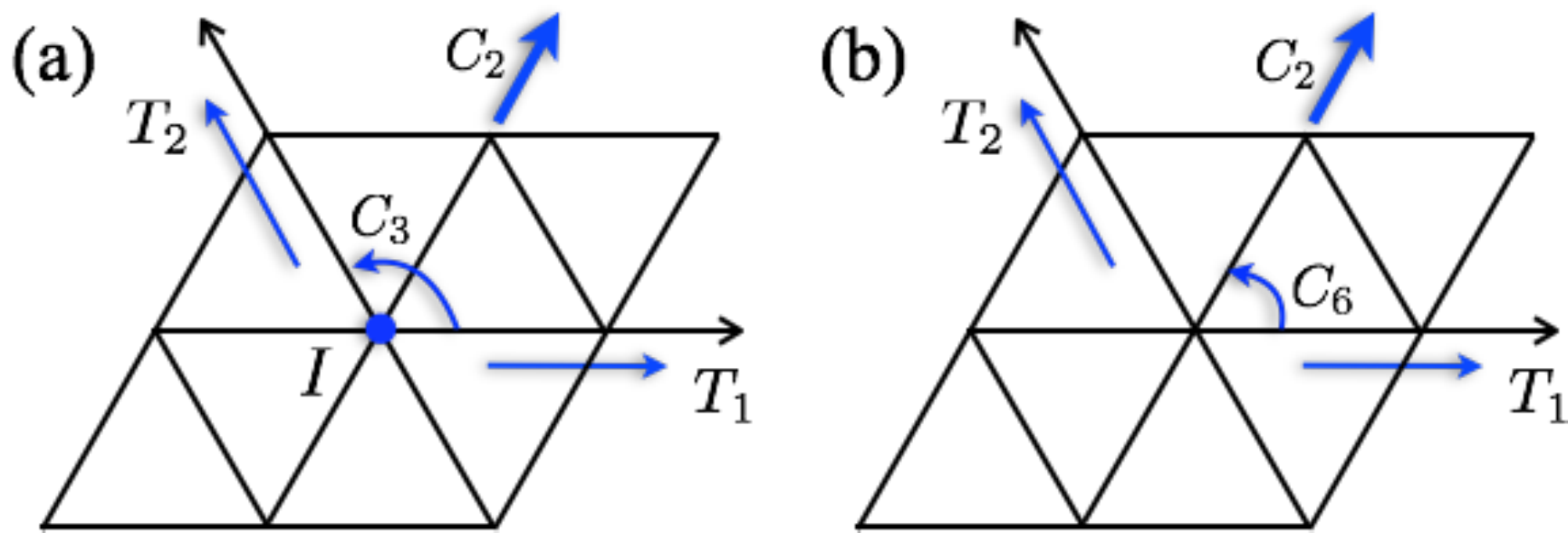}
\caption{(a) The intralayer symmetries of the R$\bar{3}$m
space group for YbMgGaO$_4$~\cite{YaodongPRB}.
(b) The same lattice symmetry group with a different 
complete set of elementary transformations. Here
 ${S_6 \equiv C_3^{-1}I}$. The bold arrow is the axis 
 for the $C_2$ rotation (see Appendix).}
\label{fig1}
\end{figure}

Apart from the absence of magnetic ordering, the heat capacity was found
to be ${C_v \propto T^{0.7}}$ at low temperatures~\cite{srep,YueshengPRL,Martin2016,Shiyan2016},
and is close to the well-known $T^{2/3}$ heat 
capacity~\cite{Motrunich2005,LeeLeePRL,Patrick1992}. 
The latter was the one obtained within a random phase 
approximation for the spinon-gauge coupling in a 
spinon Fermi surface U(1) QSL~\cite{Motrunich2005,LeeLeePRL,Patrick1992}.
More substantially, the broad continuum~\cite{YaoShenNature,Martin2016}
of the magnetic excitation with a clear dispersion for the upper
excitation edge agrees reasonably with the particle-hole continuum
of the spinon Fermi surface~\cite{YaoShenNature}. 
However, due to the scattering with the phonon degrees of freedom,
the thermal transport measurement in YbMgGaO$_4$ was unable to 
extract the intrinsic magnetic contribution to the thermal
conductivity~\cite{Shiyan2016}. Partly motivated by the spin 
liquid behaviors in YbMgGaO$_4$ and more broadly 
by the families of rare-earth magnets with identical 
structures, in this paper, we carry out a systematic 
projective symmetry group (PSG) analysis for a triangular 
lattice Mott insulator with spin-orbital-entangled local 
moments. Unlike the cases for the spin-only moments in
the pioneering work by X.-G.~Wen~\cite{XGWen}, 
the space group symmetry operation, in particular, 
the rotation, transforms both the position and the 
orientation of the Yb local moments~\cite{YaodongPRB,Yaodong201608}.
We find that, among the eight U(1) QSL states, the spinon 
mean-field state that was introduced in Ref.~\onlinecite{YaoShenNature} 
and labeled as the U1A00 state in our PSG classification, 
contains a large spinon Fermi surface and gives a large 
spinon scattering density of states that is consistent 
with the inelastic neutron scattering (INS) results.

The following part of the paper is organized as follows.
In Sec.~\ref{sec2}, we describe the space group symmetry and the 
the multiplication rules for the symmetry transformation. 
In Sec.~\ref{sec3}, we introduce the fermionic spinon 
construction and the fermionic spinon mean-field Hamiltonian. 
In Sec.~\ref{sec4}, we explain the scheme for the projective symmetry 
group classification when the spin-orbit coupling is present. 
In Sec.~\ref{sec5}, we explain the relationship between the 
spinon band structure and the projective symmetry group of the 
spinon mean-field states.
In Sec.~\ref{sec6}, we focus on the U1A00 state and study the 
spectroscopic properties of this state.
Finally in Sec.~\ref{sec7}, we discuss the experimental relevance
and remark on the thermal transport result and the competing 
scenarios and proposals. The details of the calculation are 
presented in the Appendices.

\section{Space group symmetry}
\label{sec2}

It was pointed out that the intralayer 
symmetries involves two translations, $T_1$ and $T_2$, one two-fold 
rotation, $C_2$, one three-fold rotation, $C_3$, and one spatial 
inversion $I$ (see Fig.~\ref{fig1}(a))~\cite{YaodongPRB,Yaodong201608}.
Here we use a different complete set of elementary
transformations for the space group symmetries that involve
two translations, $T_1$ and $T_2$, one two-fold rotation, $C_2$,
and one more operation, $S_6$ (see the definition in Fig.~\ref{fig1}(b)).
It is ready to confirm ${I = S_6^3, C_3^{} = S_6^2}$ with the 
definition ${S_6^{} \equiv C_3^{-1} I}$. The multiplication rules of 
this symmetry group is given as
\begin{eqnarray}
T_1^{-1} T_2^{} T_1^{} T_2^{-1} &=&
T_1^{-1} T_2^{-1} T_1^{} T_2^{}  = 1 ,
\label{eq1}
\\
C_2^{-1} T_1^{} C_2^{} T_2^{-1} &=&
C_2^{-1} T_2^{} C_2^{} T_1^{-1}  = 1 ,
\label{eq2}
\\
S_6^{-1} T_1^{} S_6^{} T_2^{}   &=&
S_6^{-1} T_2^{} S_6^{} T_2^{-1} T_1^{-1} =1 ,
\label{eq3}
\\
(C_2^{})^2 = (S_6^{})^6 &=&
(S_6^{} C_2^{})^2 = 1.
\label{eq4}
\end{eqnarray}
Due to the presence of time reversal in YbMgGaO$_4$~\cite{YueshengPRL,YueshengmuSR,YaoShenNature,Martin2016},
we further supplement the symmetry group with 
the time reversal ${\mathcal T}$ such that
${{\mathcal O}^{-1}{\mathcal T} {\mathcal O} {\mathcal T} = 1}$
and ${\mathcal{T}^2 = 1}$, where $\mathcal{O}$ is a lattice 
symmetry operation. 

\begin{table}
\renewcommand\arraystretch{1.4}
\begin{tabular}{ccccc}
\hline\hline
U(1) QSL & $W^{T_1}_{\boldsymbol r}$ & $W^{T_2}_{\boldsymbol r}$
& $W^{C_2}_{\boldsymbol r}$ & $W^{S_6}_{\boldsymbol r}$
\\
\hline
U1A00 & $I_{2\times 2}^{}$ & $I_{2\times 2}^{}$ & $I_{2\times 2}^{}$ & $I_{2\times 2}^{}$
\\
U1A10 & $I_{2\times 2}^{}$ & $I_{2\times 2}^{}$ & $i\sigma^y$ & $I_{2\times 2}^{}$
\\
U1A01 & $I_{2\times 2}^{}$ & $I_{2\times 2}^{}$ & $I_{2\times 2}^{}$ & $i \sigma^y $
\\
U1A11 & $I_{2\times 2}^{}$ & $I_{2\times 2}^{}$ & $i\sigma^y$   &     $i \sigma^y $
\\
\hline\hline
\end{tabular}
\caption{List of the gauge transformations for the 
four U1A PSGs. For the time reversal, all PSGs here have  
${W_{\boldsymbol{r}}^{\mathcal{T}} = I_{2\times 2}}$.
The last two letters in the labels of the U(1) QSLs
are extra quantum numbers in the PSG classification~\cite{Supple}. }
\label{tab1}
\end{table}

\section{Fermionic parton construction}
\label{sec3}

To describe the U(1) QSL that 
we propose for YbMgGaO$_4$, we introduce the fermionic spinon operator
$f_{\boldsymbol{r}\alpha}({\alpha=\uparrow,\downarrow}$)
that carries spin-1/2, and express the Yb local moment
as 
\begin{equation}
{\boldsymbol{S}^{}_{\boldsymbol{r}}} =
\frac{1}{2} \sum_{\alpha,\beta}
f^{\dagger}_{\boldsymbol{r}\alpha}
\boldsymbol{\sigma}^{}_{\alpha\beta} f^{}_{{\boldsymbol{r}}\beta},
\end{equation}
where ${\boldsymbol{\sigma}=(\sigma^x, \sigma^y, \sigma^z)}$
is a vector of Pauli matrices. We further impose a constraint
$\sum_{\alpha} f^{\dagger}_{\boldsymbol{r}\alpha}
f^{}_{\boldsymbol{r}\alpha} =  1$
on each site to project back to the physical Hilbert space of the spins.
The choice of fermionic spinons allows a local SU(2) gauge freedom~\cite{XGWen}.

As a direct consequence of the spin-orbital entanglement, the spinon
mean-field Hamiltonian for the U(1) QSL should generically involve
both spin-preserving and spin-flipping hoppings, and
has the following form
\begin{eqnarray}
H_{\text{MF}}^{} = -
\sum_{ ({\boldsymbol r}{\boldsymbol r}')} \sum_{ \alpha\beta } \big[\,
t_{{\boldsymbol r}{\boldsymbol r}',\alpha\beta}^{}
f^{\dagger}_{{\boldsymbol r} \alpha} f^{}_{\boldsymbol{r}'\beta} + h.c.
\,\big],
\label{eqmft}
\end{eqnarray}
where $t_{{\boldsymbol r}{\boldsymbol r}',\alpha\beta}$ is the
spin-dependent hopping. The choice of the mean-field ansatz in
Eq.~(\ref{eqmft}) breaks the local SU(2) gauge freedom down to U(1).
Here, to get a more compact form for Eq.~(\ref{eqmft}),
we follow Ref.~\onlinecite{Jason} and introduce the extended 
Nambu spinor representation for the spinons such that
$\Psi_{\boldsymbol{r}}^{} = (f^{}_{\boldsymbol{r}\uparrow},
f^{\dagger}_{\boldsymbol{r}\downarrow}, f^{}_{\boldsymbol{r}\downarrow},
- f^{\dagger}_{\boldsymbol{r}\uparrow})^T$ and
\begin{eqnarray}
H_{\text{MF}}^{} = - \frac{1}{2}
\sum_{(\boldsymbol{r},\boldsymbol{r}')}
\big[
\Psi_{\boldsymbol{r}}^{\dagger}
u_{\boldsymbol{r} \boldsymbol{r}'}^{}
\Psi^{}_{\boldsymbol{r}'}
+ h.c. \big],
\end{eqnarray}
where $u_{\boldsymbol{r} \boldsymbol{r}'}^{}$ is a hopping matrix 
that is related to $t_{{\boldsymbol r}{\boldsymbol r}',\alpha\beta}$.
With the extended Nambu spinor, the spin operator
$\boldsymbol{S}_{\boldsymbol{r}}$ and the generator 
$\boldsymbol{G}_{\boldsymbol{r}}$ for the SU(2) gauge 
transformation are given by~\cite{XGWen,Hermele,HermeleHoneycomb,PhysRevB.93.094437,PhysRevB.86.085145}
\begin{eqnarray}
\boldsymbol{S}_{\boldsymbol{r}} &=&  \frac{1}{4} \Psi^\dagger_{\boldsymbol{r}}
( { \boldsymbol{\sigma}  \otimes  I_{2\times 2}} ) \Psi_{\boldsymbol{r}}^{}, 
\\
\boldsymbol{G}_{\boldsymbol{r}} &=& \frac{1}{4} \Psi^\dagger_{\boldsymbol{r}}
( {I_{2\times 2}}  \otimes {\boldsymbol{\sigma}} ) \Psi_{\boldsymbol{r}}^{},
\label{gaugeg}
\end{eqnarray}
where $I_{2\times 2}$ is a ${2\times 2}$ identity matrix.
Under the symmetry operation $\mathcal{O}$, $\Psi_{\boldsymbol{r}}^{}$
transforms as
\begin{equation}
\Psi_{\boldsymbol{r}}^{} \rightarrow
\mathcal{U}_{\mathcal{O}}^{}
\mathcal{G}^{\mathcal{O}}_{\mathcal{O} (\boldsymbol{r})}
\Psi_{\mathcal{O}(\boldsymbol{r})}^{} =
\mathcal{G}^{\mathcal{O}}_{\mathcal{O} (\boldsymbol{r})}
{\mathcal U}_{\mathcal{O}}^{}
\Psi_{\mathcal{O}(\boldsymbol{r})}^{} ,
\label{op}
\end{equation}
where $\mathcal{G}^{\mathcal{O}}_{\mathcal{O} (\boldsymbol{r})}$ is the local gauge
transformation that corresponds to the symmetry operation $\mathcal{O}$,
and we add a spin rotation ${\mathcal U}_{\mathcal{O}}^{}$ because
the spin components are transformed when $\mathcal{O}$ involves a rotation.
In Eq.~(\ref{op}), the gauge transformation and the spin rotation
are commutative~\cite{Kim} simply because
$[S^{\mu}_{\boldsymbol r}, G^{\nu}_{\boldsymbol r}] = 0$.
Moreover, from Eq.~(\ref{gaugeg}),
the gauge transformation $\mathcal{G}^{\mathcal{O}}_{\boldsymbol{r}}$
is block diagonal with $\mathcal{G}_{{\boldsymbol r}}^{\mathcal O}
= I_{2 \times 2} \otimes W_{{\boldsymbol r}}^{\mathcal O}$,
where $W_{{\boldsymbol r}}^{\mathcal O}$ is a $2\times 2$ matrix (see Appendix).

\section{Projective symmetry group classification}
\label{sec4}

For the spinon mean-field Hamiltonian in Eq.~(\ref{eqmft}), the lattice
symmetries are realized projectively and form the projective
symmetry group (PSG). To respect the lattice symmetry 
transformation $\mathcal{O}$, the mean-field ansatz should satisfy
\begin{eqnarray}
u_{{\boldsymbol r}{\boldsymbol r}'}^{}
= {\mathcal G}^{\mathcal{O}\dagger}_{\mathcal{O}(\boldsymbol{r})}
{\mathcal U}^\dagger_{\mathcal{O}}
u_{\mathcal{O}(\boldsymbol{r})\mathcal{O}(\boldsymbol{r}')}^{}
{\mathcal U}^{}_{\mathcal{O}}
{\mathcal G}^{\mathcal{O}}_{\mathcal{O}(\boldsymbol{r}')}.
\end{eqnarray}
The ansatz itself is invariant under the so-called invariant gauge
group (IGG) with ${u_{{\boldsymbol r}{\boldsymbol r}'}^{}
= {\mathcal G}^{1\dagger}_{\boldsymbol{r}}
u_{{\boldsymbol r}{\boldsymbol r}'}^{}
{\mathcal G}^{1}_{\boldsymbol{r}'}}$.
The IGG can be regarded as a set of gauge transformations
that correspond to the identity transformation.
For an U(1) QSL, IGG = U(1).

A general group relation $\mathcal{O}_1 \mathcal{O}_2
\mathcal{O}_3 \mathcal{O}_4 = 1$ for the lattice symmetry
turns into the following group relation for the PSG
\begin{eqnarray}
&&{\mathcal U}^{}_{\mathcal{O}_1} {\mathcal G}^{\mathcal{O}_1}_{\boldsymbol{r}}
{\mathcal U}^{}_{\mathcal{O}_2}
{\mathcal G}^{\mathcal{O}_2}_{\mathcal{O}_2\mathcal{O}_3\mathcal{O}_4(\boldsymbol{r})}
{\mathcal U}^{}_{\mathcal{O}_3}
{\mathcal G}^{\mathcal{O}_3}_{\mathcal{O}_3\mathcal{O}_4(\boldsymbol{r})}
{\mathcal U}^{}_{\mathcal{O}_4}
{\mathcal G}^{\mathcal{O}_4}_{\mathcal{O}_4(\boldsymbol{r})}
\nonumber
\\
&=& {\mathcal U}^{}_{\mathcal{O}_1}{\mathcal U}^{}_{\mathcal{O}_2}
{\mathcal U}^{}_{\mathcal{O}_3}{\mathcal U}^{}_{\mathcal{O}_4}
{\mathcal G}^{\mathcal{O}_1}_{\boldsymbol{r}}
{\mathcal G}^{\mathcal{O}_2}_{\mathcal{O}_2\mathcal{O}_3\mathcal{O}_4(\boldsymbol{r})}
{\mathcal G}^{\mathcal{O}_3}_{\mathcal{O}_3\mathcal{O}_4(\boldsymbol{r})}
{\mathcal G}^{\mathcal{O}_4}_{\mathcal{O}_4(\boldsymbol{r})}
\label{eq15}
\\
 &\in & \text{IGG},
\end{eqnarray}
where we used the fact that the gauge transformation commutes with the
spin rotation. As the series of rotations
$ {\mathcal{O}_1} {\mathcal{O}_2} {\mathcal{O}_3} {\mathcal{O}_4} $
either rotate the spinons by $0$ or $2\pi$,
\begin{eqnarray}
{{\mathcal U}^{}_{\mathcal{O}_1}{\mathcal U}^{}_{\mathcal{O}_2}
{\mathcal U}^{}_{\mathcal{O}_3}{\mathcal U}^{}_{\mathcal{O}_4}
= \pm I_{4 \times 4}},
\end{eqnarray}
where $I_{4 \times 4}$ is a ${4\times 4}$
identity matrix. Since ${\{\pm I_{4 \times 4}\} \subset \text{IGG}}$,
then 
\begin{eqnarray}
{{\mathcal G}^{\mathcal{O}_1}_{\boldsymbol{r}}
{\mathcal G}^{\mathcal{O}_2}_{\mathcal{O}_2\mathcal{O}_3\mathcal{O}_4(\boldsymbol{r})}
{\mathcal G}^{\mathcal{O}_3}_{\mathcal{O}_3\mathcal{O}_4(\boldsymbol{r})}
{\mathcal G}^{\mathcal{O}_4}_{\mathcal{O}_4(\boldsymbol{r})} \in \text{IGG}}.
\end{eqnarray}
This immediately indicates that, to classify the PSGs for a spin-orbit-coupled 
Mott insulator, we only need to focus on the gauge part, first find the gauge transformation
with the same procedures as those for the conventional Mott insulators
with spin-only moments~\cite{XGWen}, and then account for the spin rotation.

For the mean-field ansatz in $H_{\text{MF}}^{}$, we choose the 
``canonical gauge'' for the IGG with 
\begin{eqnarray}
\text{IGG} = {\{ I_{2\times 2} \otimes
e^{i \phi \sigma^z}| \phi \in [0,2\pi) \}}. 
\end{eqnarray}
Under the canonical gauge, the gauge transformation associated with
the symmetry operation $\mathcal{O}$ takes the form of
\begin{eqnarray}
\mathcal{G}^{\mathcal O}_{\boldsymbol{r}} & = &
I_{2\times 2} \otimes W^{\mathcal O}_{\boldsymbol{r}} 
\nonumber \\
&\equiv & I_{2\times 2} \otimes \big[ (i \sigma^x)^{n^{}_{\mathcal O}}
e^{i \phi^{}_{\mathcal O} [{\boldsymbol r}] \sigma^z} \big],
\end{eqnarray}
where ${n_{\mathcal O}^{} = 0,1}$. For translations, one can always 
choose a gauge such that 
\begin{eqnarray}
W^{T_1}_{\boldsymbol{r}} &=& (i\sigma^x)^{n_{1}^{}},\\
W^{T_2}_{\boldsymbol{r}}  &=& (i\sigma^x)^{n_{2}^{}} e^{i\phi_{2}^{} [x,y]\sigma^z}
\end{eqnarray}  
with
$n_1^{},n_2^{}=0,1$ and $\phi_2^{} [0,y] = 0$.
The group relation in Eq.~(\ref{eq3}) further demands
$n_1^{} = n_2^{} = 0$. Thus the group relation
in Eq.~(\ref{eq1}) gives
$W^{T_1}_{\boldsymbol{r}} = 1, W^{T_2}_{\boldsymbol{r}} 
= e^{i x \phi_1^{} \sigma^z}$,
where $\phi_1^{}$ is the flux through each unit cell of the triangular
lattice and takes the value of $0 \text{ or } \pi$ (see Appendix).
The PSGs with $\phi_1^{}=0 \, (\pi)$ are labeled by U1A (U1B).
Among the sixteen algebraic PSGs that we find, eight
unphysical solutions have ${\mathcal{T}^2 =1}$ for the spinons
and give vanishing spinon hoppings everywhere. In
Tab.~\ref{tab1} and the Appendix, 
we list the remaining eight PSGs that have ${\mathcal{T}^2 =-1}$ consistent 
with the fact that fermionic spinons are Kramers doublets (see Appendix).

\begin{figure}[tp]
\centering
\includegraphics[width=.23\textwidth]{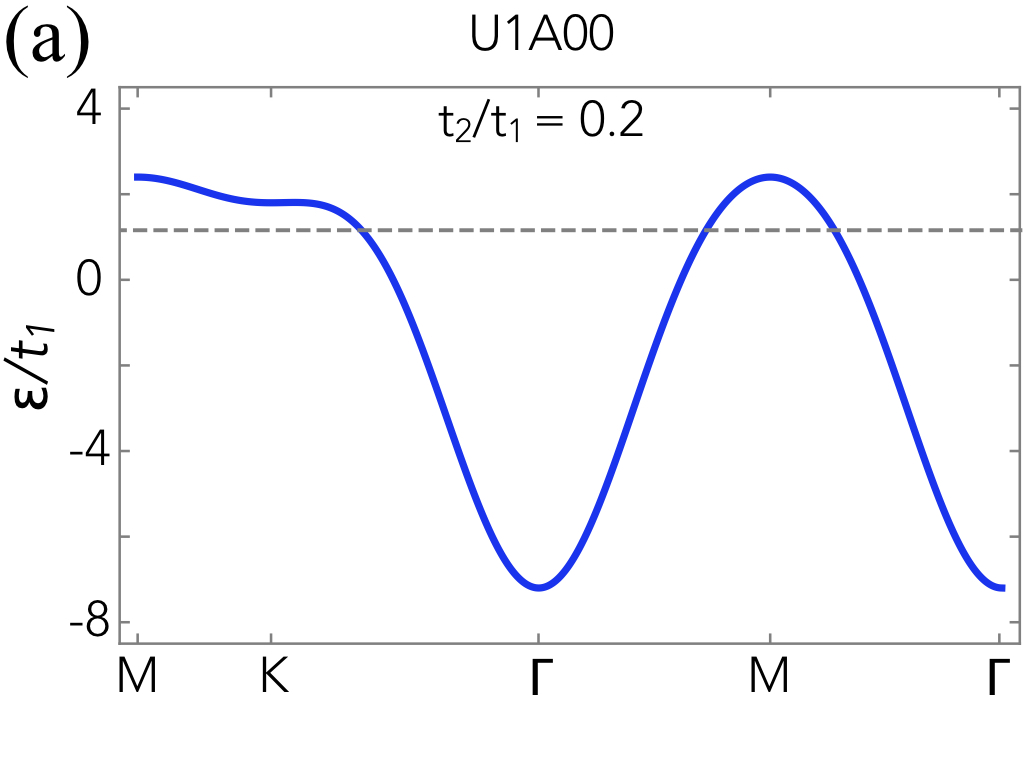}
\includegraphics[width=.23\textwidth]{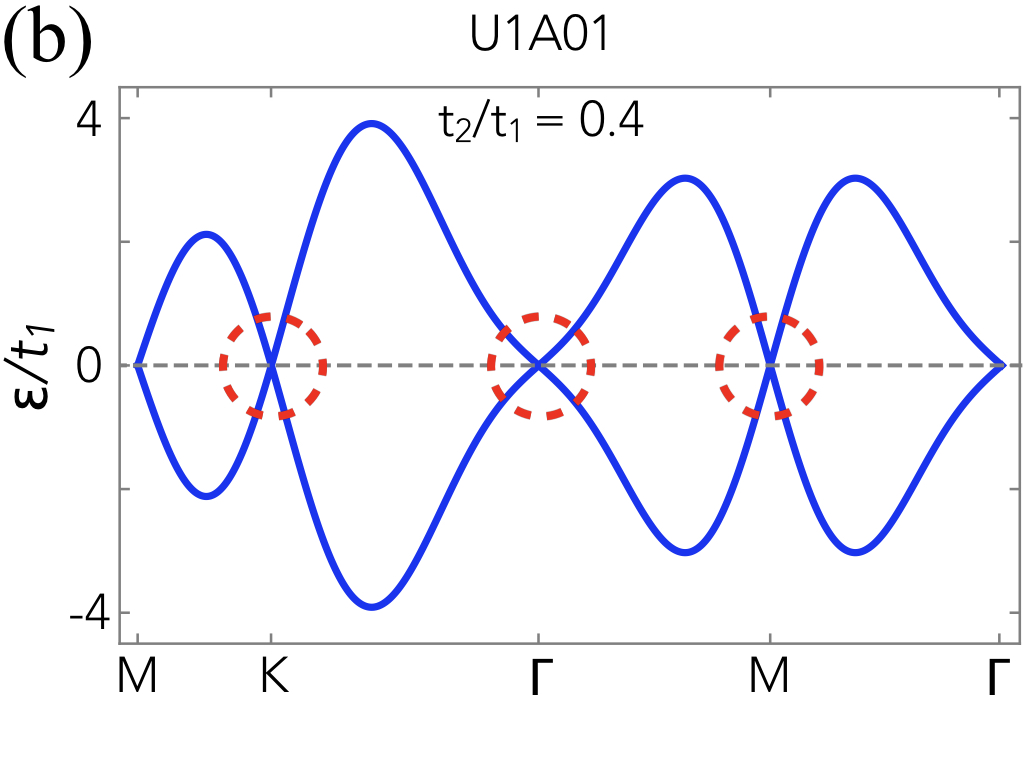}
\includegraphics[width=.23\textwidth]{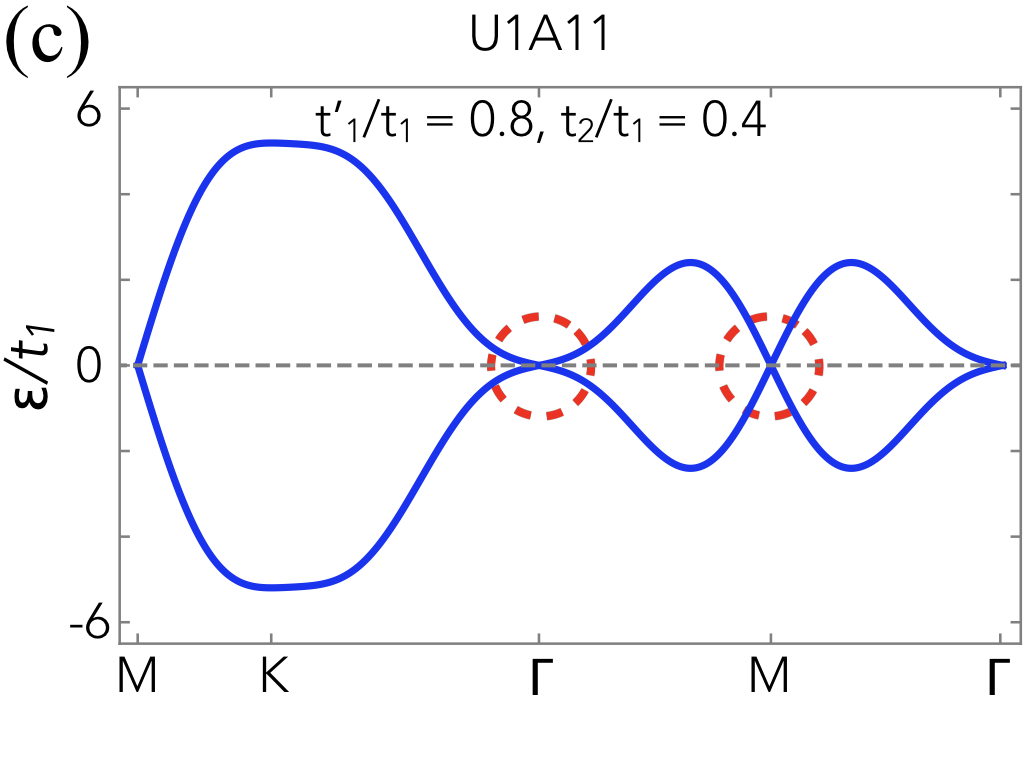}
\includegraphics[width=.23\textwidth]{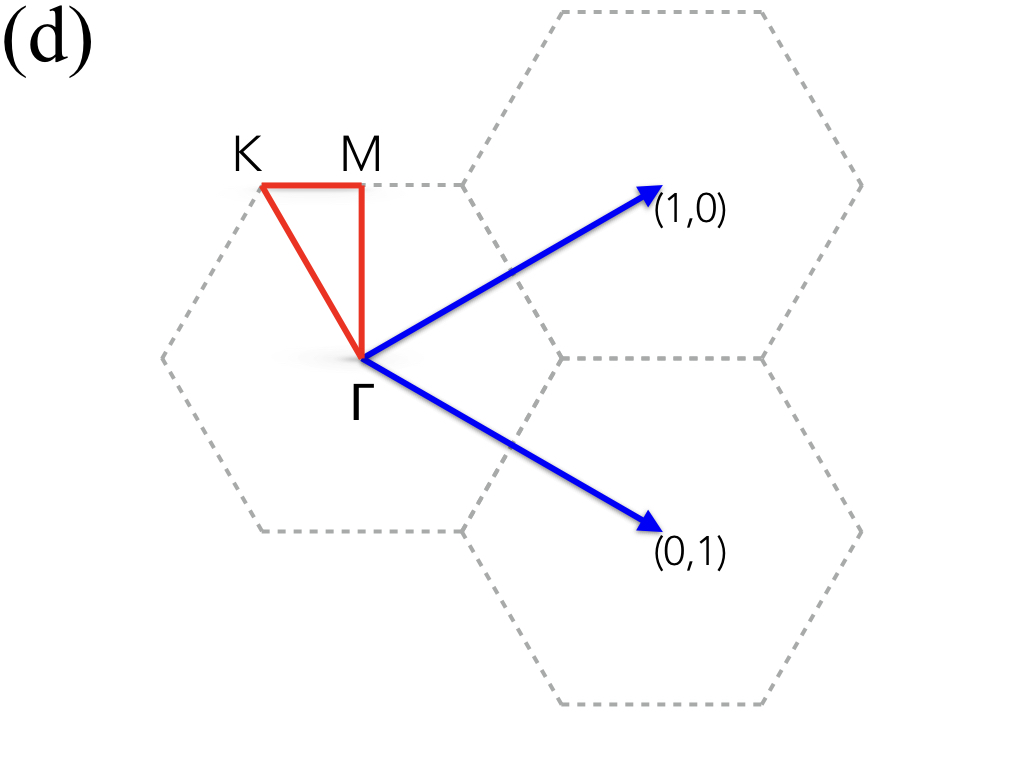}
\caption{(a,b,c) The mean-field spinon bands
along the high-symmetry momentum lines (see (d)) of the
U1A00, U1A01 and U1A11 states,
where $t_1, t_1'$ and $t_2$ are hoppings in their spinon
mean-field Hamiltonians (see Appendix). The Dirac cones are 
highlighted in dashed circles. The dashed line refers to
the Fermi level. (d) The Brioullin zone 
of the triangular lattice.}
\label{fig2}
\end{figure}

\section{Mean-field states}
\label{sec5}

Here we obtain the spinon mean-field Hamiltonian from Tab.~\ref{tab1}
and explain why the U1A00 state stands out as the candidate ground state
for YbMgGaO$_4$. We start with the U1A states. Among the four U1A states,
the U1A10 state gives a vanishing mean-field Hamiltonian for
the spinon hoppings between the first and the second neighbors,
the remaining ones except the U1A00 state all have
symmetry protected band touchings
at the spinon Fermi level (see Fig.~\ref{fig2}). 
To illustrate the idea~\cite{Lu2016}, we consider
the U1A01 state where the spinon Hamiltonian has the form
$H_{\text{MF}}^{\text{U1A01}} = \sum_{\boldsymbol{k}}
h_{\alpha\beta}^{} ({\boldsymbol k})
f^{\dagger}_{{\boldsymbol k}\alpha}
f^{}_{{\boldsymbol k}\beta}$ in the
momentum space and $h({\boldsymbol k})$
is a $2\times 2$ matrix with
\begin{eqnarray}
h ({\boldsymbol k})
= d_0 ({\boldsymbol k}) I_{2 \times 2}^{}
+ \sum_{\mu=1}^3 {d}_{\mu} ({\boldsymbol k})
{{\sigma}^{\mu}} .
\end{eqnarray}
For this band structure there are nondegenerate band touchings 
at $\Gamma$, M and K points that are protected 
by the PSG of the U1A01 state. Under the operation $S_6$, the PSG demands 
that
spinons to transform as
\begin{eqnarray}
f^{}_{\boldsymbol{k}\uparrow} & \rightarrow & - e^{-i{\pi}/{3}}
f^{\dagger}_{-S_6^{-1} \boldsymbol{k},\downarrow } , \\
f^{}_{\boldsymbol{k}\downarrow} & \rightarrow &
e^{i{\pi}/{3}} f^{\dagger}_{{-S_6^{-1} \boldsymbol{k}},\uparrow }.
\end{eqnarray}
Applying $S_6$ three times and keeping $H_{\text{MF}}$ invariant, 
we require 
\begin{eqnarray}
{h({\boldsymbol k}) = - [\sigma^y h({\boldsymbol k}) \sigma^y ]^{T} }
\end{eqnarray}
which forces ${d_0({\boldsymbol k}) = 0}$. The time reversal symmetry
(${\mathcal{T} = i\sigma^y \otimes I_{2\times 2} K}$) further requires
that ${d_{\mu} ({\boldsymbol k}) = - d_{\mu} (-{\boldsymbol k})}$.
Thus we have symmetry protected band touchings with $h({\boldsymbol k}) = 0$
at the time reversal invariant momenta $\Gamma$ and M.
The K points are invariant under $C_2$ and $S_6$ because 
the spinon partile-hole transformation is involved for $S_6$ (see Appendix). 
Using those two symmetries, we further establish the band touching at the K points.
Likewise, for the U1A11 state, the PSG demands the band touchings at $\Gamma$
and M points. Because there are only two spinon bands for the U1A states,
these band touchings generically occur at the spinon Fermi level.

Due to the Dirac band touchings at the Fermi level, the low-energy dynamic spin
structure factor, that measures the spinon particle-hole continuum, is concentrated
at a few discrete momenta that correspond to the intra-Dirac-cone and the
inter-Dirac-cone scatterings~\cite{YaoShenNature}. Clearly, this is inconsistent 
with the recent INS result that observes a broad continuum 
covering a rather large portion of the Brillouin zone~\cite{YaoShenNature,Martin2016}.

For the U1B states, the spinons experience a $\pi$ background flux 
in each unit cell. The direct consequence of the $\pi$ 
background flux is that the U1B states support an enhanced periodicty of the
dynamic spin structure in the Brillouin zone~\cite{Wen2002175,XGWen,Hermele2}.
Such an enhanced periodicity is absent in the INS 
result~\cite{YaoShenNature,Martin2016}. In particular, unlike what one 
would expect for an enhanced periodicity, the spectral
intensity at the $\Gamma$ point is drastically different
from the one at the M point in the existing
experiments~\cite{YaoShenNature,Martin2016}.

The above analysis leads to the conclusion that the U1A00 state
is the most promising candidate U(1) QSL for YbMgGaO$_4$,
and this conclusion is {\it independent} from any microscopic model.
The spinon mean field Hamiltonian, allowed by the U1A00 PSG,
is remarkably simple and is given as~\footnote{In the previous
work~\cite{YaoShenNature}, only the nearest-neighbor spinon 
hopping is included.}

\begin{equation}
H_{\text{MF}}^{\text{U1A00}} = -t_1 
\sum_{\langle \boldsymbol{r} \boldsymbol{r}' \rangle, \alpha }
f^{\dagger}_{\boldsymbol{r}\alpha} f^{}_{\boldsymbol{r}\alpha}
-t_2 
\sum_{\langle\langle \boldsymbol{r} \boldsymbol{r}' \rangle\rangle, \alpha}
f^{\dagger}_{\boldsymbol{r}\alpha} f^{}_{\boldsymbol{r}\alpha} ,
\label{t1t2eq1}
\end{equation}
where the spinon hopping is isotropic for the first and the second neighbors.
This mean-field state only has a single band that is 1/2-filled, 
so it has a large spinon Fermi surface. From $H_{\text{MF}}^{\text{U1A00}}$, 
we construct the mean-field ground state by filling the spinon Fermi sea,
\begin{equation}
{{|\Psi_{\text{MF}}^{\text{U1A00}} \rangle} =
\prod_{ \epsilon_{\boldsymbol{k}}^{} < \epsilon_{\text F}^{} }
f^\dagger_{ {\boldsymbol k} \uparrow}
f^\dagger_{ {\boldsymbol k} \downarrow}
\, {|0\rangle}}
\end{equation}
where $\epsilon_{\boldsymbol{k}}^{}$
is the spinon dispersion and $\epsilon_{\text F}^{}$ is the spinon
Fermi energy. The mean-field variational energy is
\begin{eqnarray}
E_{\text{var}}^{} = \langle \Psi_{\text{MF}}^{\text{U1A00}}|
H_{\text{spin}}^{} | \Psi_{\text{MF}}^{\text{U1A00}} \rangle ,
\end{eqnarray}
where 
\begin{eqnarray}
H_{\text{spin}} &=& \sum_{\langle {\boldsymbol r} {\boldsymbol r}' \rangle}
J_{zz}^{} S^z_{\boldsymbol r} S^z_{{\boldsymbol r}'}
+{ J_{\pm}^{} ( S^+_{\boldsymbol r} S^-_{{\boldsymbol r}'}
+  S^-_{\boldsymbol r} S^+_{{\boldsymbol r}'})} 
\nonumber \\
&&  \quad + {J_{\pm\pm}^{} (\gamma_{{\boldsymbol r} {\boldsymbol r}' }^{}
S^+_{\boldsymbol{r}} S^+_{\boldsymbol{r}'} +
\gamma_{{\boldsymbol r} {\boldsymbol r}' }^{\ast}
S^-_{\boldsymbol{r}} S^-_{\boldsymbol{r}'} )} 
\nonumber \\
&&
\quad {- \frac{i}{2} J_{z\pm}^{} \big[  (\gamma^{\ast}_{{\boldsymbol r}{\boldsymbol r}'}
S^+_{\boldsymbol r} - \gamma^{}_{{\boldsymbol r}{\boldsymbol r}'} S^-_{\boldsymbol r})} 
S_{\boldsymbol{r}'}^z
\nonumber \\
&& 
\quad\quad\quad\quad + S^z_{\boldsymbol{r}} (\gamma^{\ast}_{{\boldsymbol r}{\boldsymbol r}' }
S^+_{{\boldsymbol r}'} -\gamma^{}_{{\boldsymbol r}{\boldsymbol r}' }
S^-_{{\boldsymbol r}'} ) \big]
\end{eqnarray}
is the microscopic spin model that was introduced
in Refs.~\onlinecite{YueshengPRL,YaodongPRB},
and $\gamma_{{\boldsymbol r}{\boldsymbol r}'}^{}$ is a
bond-dependent phase factor due to the spin-orbit-entangled
nature of the Yb moments~\cite{YaodongPRB}.
The anisotropic nature of the spin interaction
has been clearly supported by the recent
polarized neutron scattering measurement~\cite{1705}. 
For the specific choice with
${J_{\pm} = 0.915 J_{zz}}$,
we find the minimum variational energy ${E_{\text{var}}
= -0.39 J_{zz}}$ and occurs at ${t_2 = 0.2 t_1}$ (see Appendix). 
Here, the expectation values of the $J_{\pm\pm}$ and $J_{z\pm}$ 
interactions simply vanish, and this is an artifact 
of the free spinon mean-field theory with the isotropic hoppings 
in Eq.~(\ref{t1t2eq1}). We here establish that the U1A00 state
is a spinon Fermi surface U(1) QSL.

\section{Spectroscopic properties}
\label{sec6}

For the U1A00 state, the dynamic spin structure essentially detects 
the spinon particle-hole excitation across the Fermi surface. 
The information about the Fermi surface is encoded in 
the profile of the dynamic spin structure factor.  
We evaluate the dynamic spin structure factor within the 
free spinon mean-field theory (see Appendix) 
(see Fig.~\ref{fig3}(a)). Qualitatively similar to 
the mean-field theory with only first neighbor spinon hoppings, 
the improved free-spinon mean-field theory 
of $H_{\text{MF}}^{\text{U1A00}}$ captures 
the crucial features of the INS results~\cite{YaoShenNature,Martin2016}.
The spinon particle-hole continuum covers 
a large portion of the Brillouin zone,
and vanishes beyond the spinon bandwidth. 
More importantly, the ``V-shape'' upper excitation 
edge near the $\Gamma$ point in Fig.~\ref{fig3}(a) was 
clearly observed in the experiments~\cite{YaoShenNature,Martin2016},
and the slope of the ``V-shape'' is the Fermi velocity. 

\begin{figure}[tp]
\centering
\includegraphics[width=.238\textwidth]{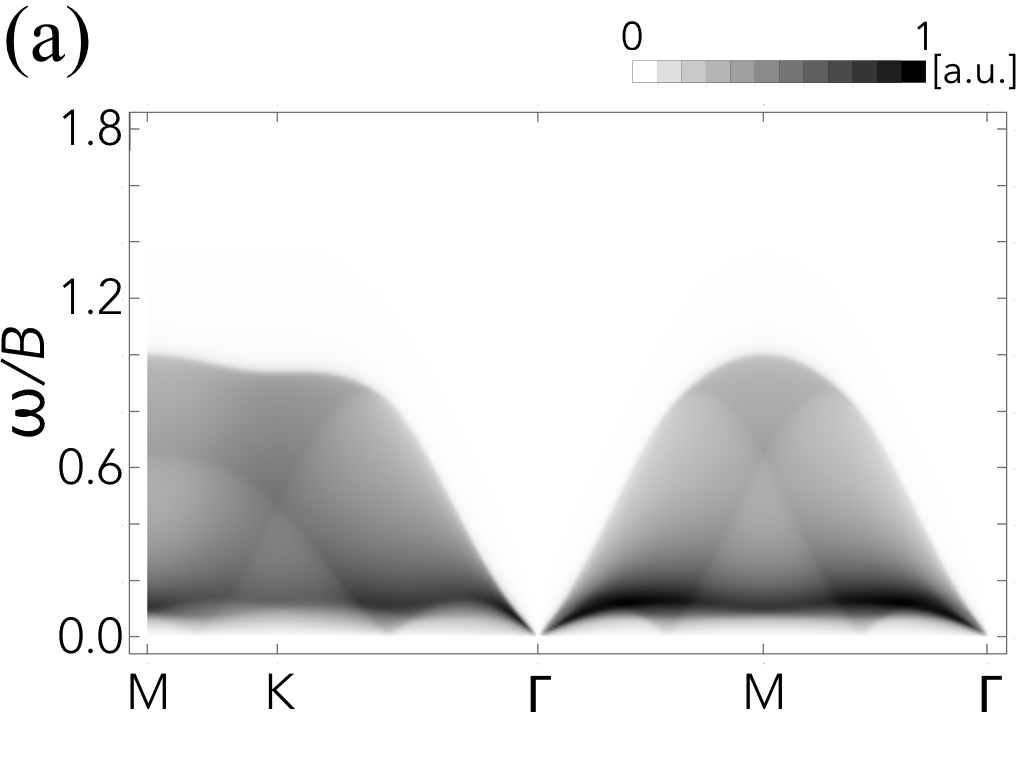}
\includegraphics[width=.238\textwidth]{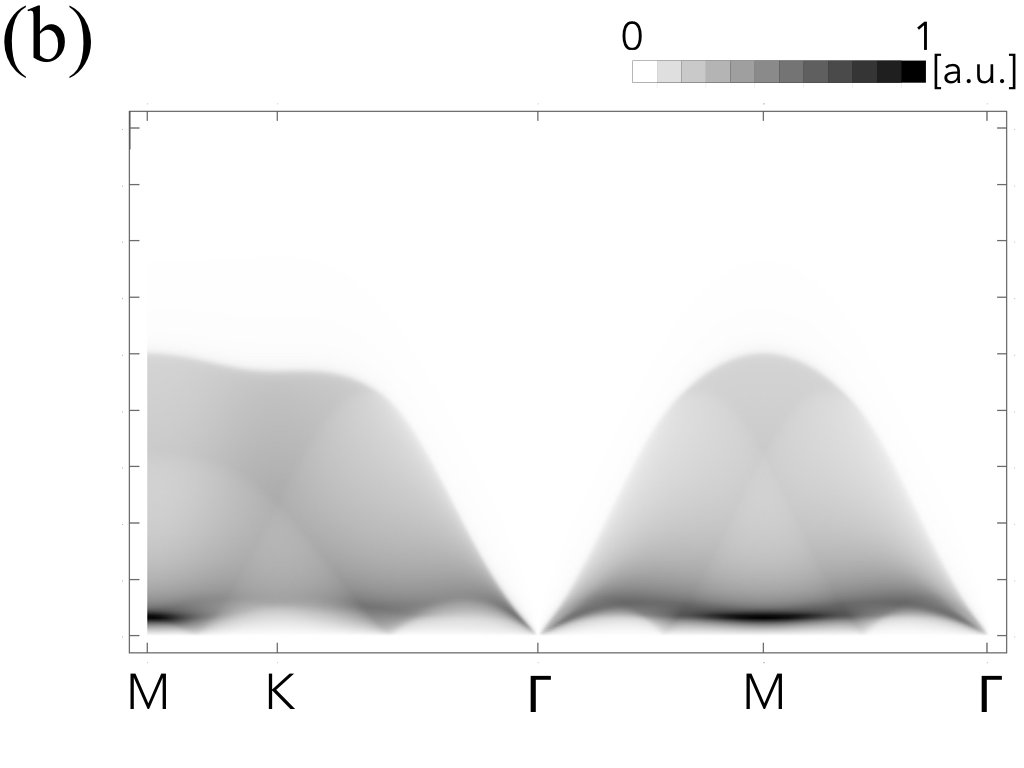}
\caption{(a) ${\mathcal S}({\boldsymbol q},\omega)$ 
along the high-symmetry momentum lines from 
$H_{\text{MF}}^{\text{U1A00}}$ with ${t_2 = 0.2t_1}$. 
The spinon bandwidth ${B=9.6t_1}$.
(b) The RPA corrected ${\mathcal S}^{\text{RPA}}
({\boldsymbol q},\omega)$ along the high symmetry 
momentum lines. We have set the parameters in the 
spin model to be ${J_{\pm}/J_{zz} = 0.915}$,
${J_{\pm\pm}/J_{zz} = 0.35}$, and ${J_{z\pm}/J_{zz}=0.2}$. 
The ratio $J_{zz}/t_1$ is obtained 
from Refs.~\onlinecite{YaoShenNature,YueshengPRL} and 
fixed to be $1.0$ for concreteness.}
\label{fig3}
\end{figure}

Due to the isotropic spinon hoppings, $H_{\text{MF}}^{\text{U1A00}}$
does not explicitly reflect the absence of spin-rotational symmetry 
that is brought by the $J_{\pm\pm}$ and $J_{z\pm}$ interactions. 
To incorporate the $J_{\pm\pm}$ and $J_{z\pm}$ interactions, 
we follow the phenomenological RPA treatment for the ``$t$-$J$'' 
model in the context of cuprate superconductors~\cite{PhysRevLett.82.2915}
and consider 
\begin{equation}
H = H_{\text{MF}}^{\text{U1A00}} + H_{\text{spin}}',
\end{equation}
where $H_{\text{spin}}'$ are the 
$J_{\pm\pm}$ and $J_{z\pm}$ interactions (see Appendix). 
While the free spinon results from $H_{\text{MF}}^{\text{U1A00}}$ already capture 
the main features of the neutron scattering data~\cite{YaoShenNature,Martin2016}, 
the anisotropic spin interaction $H_{\text{spin}}'$, included by RPA, 
merely redistributes the spectral weight in the momentum space. 
We find in Fig.~\ref{fig3}(b) that, the low-energy spectral weight 
at M is slightly enhanced, a feature observed in 
Refs.~\onlinecite{YaoShenNature, Martin2016}. 
From our choice of the parameters, it is plausible that this peak 
results from the proximity to a phase with a stripe-like magnetic 
order~\cite{YaoShenNature,YaodongPRB,Yaodong201608}.

\section{Discussion}
\label{sec7}

We have demonstrated that the spinon Fermi surface U(1) QSL 
gives a consistent explanation of the INS result in YbMgGaO$_4$.
Moreover, the anisotropic spin interaction, slightly enhances the 
spectral weight at the M points. The U(1) gauge fluctuation in 
the spinon Fermi surface U(1) QSL~\cite{LeeLeePRL,Motrunich2005}  
was suggested to be the cause for the sublinear temperature
dependence of the heat capacity in 
YbMgGaO$_4$~\cite{YaoShenNature,Patrick1992,YaodongPRB,Yaodong201608}.

In YbMgGaO$_4$, the coupling between the Yb moments
is relatively weak~\cite{YueshengPRL}. It is feasible to fully 
polarize the spin with experimentally accessible magnetic 
fields~\cite{YaodongPRB,Martin2016,Yaodong201608,yuesheng1702}
and to study the evolution of the magnetic properties under 
the magnetic field. Recently, two of us have predicted the 
spectral weight shift of the INS for YbMgGaO$_4$ under 
a weak magnetic field~\cite{Yaodong201703},
and the predicted spectral crossing at the $\Gamma$ point and the 
dispersion of the spinon continuum have actually been confirmed 
in the recent INS measurement~\cite{JunZhaounpub}. 
Numerically, it is useful to perform numerical calculation 
with fixed $J_{\pm}$ and $J_{zz}$ that are close to the ones 
for YbMgGaO$_4$, and obtain the phase diagram of our spin model 
by varying $J_{\pm\pm}$ and $J_{z\pm}$~\cite{YaodongPRB,Yaodong201608,ChangPRB}. 
More care needs to be paid to the disordered region of the mean-field phase 
diagram~\cite{YaodongPRB} where quantum fluctuation is found to be 
strong~\cite{YaodongPRB}. The ``$2k_{\text F}$'' oscillation in the 
spin correlation would be the strong indication of the spinon Fermi 
surface. Noteworthily recent DMRG works~\cite{PhysRevB.95.165110,1703} 
have actually provided some useful information about the ground 
states of the system, in particular, Ref.~\onlinecite{1703} suggested
the scenario of exchange disorders. Certain amount of exchange disorder 
may be created by the crystal electric field disorder that stems from
the Mg/Ga mixing in the non-magnetic layers~\cite{yuesheng1702,Martin2016},
but recent polarized neutron scattering measurement did not find strong 
exchange disorder~\cite{1705}. Regardless of the 
possibilities of exchange disorders, the spin quantum number fractionalization, 
that is one of the key properties of the QSLs, could survive even with 
weak disorders. The approach and results in our present work are 
phenomenologically based and are independent of the microscopic 
mechanism for the possible QSL ground state in YbMgGaO$_4$.  


Ref.~\onlinecite{Shiyan2016} claimed the absence of the magnetic thermal 
conductivity in YbMgGaO$_4$ by extrapolating the low-temperature thermal 
conductitivity data in the zero magnetic field. Here, we provide an 
alternative understanding for this thermal transport result. 
The hint lies in the field dependence of the thermal conductivity. 
It was found that, when strong magnetic fields are applied to YbMgGaO$_4$,
the thermal conductivity $\kappa_{xx}/T$ at 0.2K is increased compared
with the one at zero field~\cite{Shiyan2016}. 
If one ignores the disorder effect and assumes the zero-field thermal 
conductivity is a simple addition 
of the magnetic contribution and the phonon contribution with 
\begin{equation}
\kappa_{xx} = \kappa_{\text{spin},xx} + \kappa_{\text{phonon},xx},
\end{equation}
the strong magnetic field almost polarizes the spins completely 
and creates a spin gap for the magnon excitation, 
hence suppress the magnetic contribution. 
The high-field thermal conductivity would be purely given by 
the phonon contribution, and we would expect a decreasing of 
the thermal conductivity in the strong field compared to the 
zero field result. This is clearly inconsistent with the 
experimental result. Therefore, the zero-field thermal conductivity is not 
a simple addition of the magnetic contribution and the phonon contribution, 
{\sl i.e.}, 
\begin{equation}
\kappa_{xx} \neq \kappa_{\text{spin},xx} + \kappa_{\text{phonon},xx}.
\end{equation} 
This also strongly suggests {\it the presence} rather than the absence of 
magnetic excitations in the thermal conductivity result at zero magnetic field. 
If there is no magnetic excitation in the system at low temperatures, 
the low-temperature thermal conductivity at zero field should 
just be the phonon contribution, and we would expect the 
zero-field thermal conductivity to be the same as the one in the strong 
field limit, (although the intermediate field regime could be different). 
This is again inconsistent with the experiments. This means that the magnetic 
excitation certainly does not have a large gap and could just be gapless as we 
propose from the spinon Fermi surface state. In fact, the gapless nature of 
the magnetic excitation is consistent with the power-law heat capacity results
in YbMgGaO$_4$. What suppresses $\kappa_{xx}$ could arise from the mutual 
scattering between the magnetic excitations and the 
the phonons. In fact, similar field dependence of thermal conductivity 
${\kappa_{xx}}$ has been observed in other rare-earth systems such as 
Tb$_2$Ti$_2$O$_7$~\cite{Ong,XFSun,PhysRevB.87.214408}
and Pr$_2$Zr$_2$O$_7$~\cite{matsuda}. It was suggested 
there~\cite{XFSun,PhysRevB.87.214408,matsuda} that the spin-phonon 
scattering is the cause. The Yb local moment, that is a spin-orbit-entangled
object, involves the orbital degree of freedom. The orbital degree of freedom 
is sensitive to the ion position, and thus couples to the phonon strongly.
This is probably the microscopic origin for the strong
coupling between the magnetic moments and the phonons 
in the rare-earth magnets. This is quite different from 
the organic spin liquid candidates and the herbertsmithite 
kagome system where the orbital degree of freedom does not 
seem to be involved~\cite{Han2012,kappaET,organics2,organictherm}. 

If the ground state of YbMgGaO$_4$ is a QSL with the spinon Fermi 
surface, the field-driven transition from the QSL ground state to 
the fully polarized state is necessarily a
unconventional transition beyond the traditional Landau's 
paradigm and has not been studied in the previous
spin liquid candidates~\cite{Han2012,kappaET,organics2,organictherm}. 
The smooth growth of the magnetization with varying external 
fields indicates a continuous transition~\cite{YueshengPRL}. 
Since we propose YbMgGaO$_4$ to be a spinon Fermi surface 
U(1) QSL and gapless, the transition would be associated 
with the openning of the spin gap at the critical field.
The continuous nature of the transition suggests the spin gap 
to open in a continuous manner. Moreover, the spinon 
confinement would be concomitant with the spin gap that 
suppresses the spinon density of states and allows the 
instanton events of the U(1) gauge field to proliferate.
Therefore, it might be interest to identify the critical
field and obtain the critical properties of the field-driven 
transition. Thermodynamic, spectroscopic, and thermal transport 
measurements with finer field variation would be helpful. 

Finally, several families of rare-earth triangular lattice magnets 
have been discovered recently~\cite{YaodongPRB,Yaodong201608,liu2016revealing,Higuchi2016,NIENTIEDT1999478,Yamada2010,Stoyko,Cava2016}.
Their properties have not been studied carefully. Our general 
classification results and the prediction of the spectroscopic 
properties would apply to the QSL candidates 
that may emerge in these families of materials. It is certainly 
exciting if one finds the new QSL candidates in these families 
behave like YbMgGaO$_4$~\cite{YaodongPRB}.

\section{Acknowledgements}

We thank one anonymous referee
for the suggestion for improvement to this paper,
and Zhu-Xi Luo for pointing out some typos. 
G.C. acknowledges the discussion with Xuefeng Sun
from USTC and Yuji Matsuda about thermal transports 
in rare-earth magnets, and the discussion with Professor 
Sasha Chernyshev about the related matters. 
This work is supported by the Ministry of Science 
and Technology of China with the Grant 
No.2016YFA0301001 (G.C.), the Start-Up Funds of OSU 
(Y.M.L.) and Fudan University (G.C.), the National 
Science Foundation under Grant No. NSF PHY-1125915 
(Y.M.L and G.C.), the Thousand-Youth-Talent 
Program (G.C.) of China, and the first-class university
construction program of Fudan University.

\appendix

\section{The coordinate System and space group symmetry}
\label{ssec1}

Following our convention in Fig.~1 in the main text, 
we choose the coordinate system of the triangular 
lattice to be
\begin{eqnarray}
    {\boldsymbol a}_1 &=& ( 1, 0 ),  
    \\
    {\boldsymbol a}_2 &=& ( -\frac{1}{2}, \frac{\sqrt{3}}{2} ).
\end{eqnarray}

We label the triangular lattice sites by ${\boldsymbol r} 
= x\, {\boldsymbol a}_1 +  y\, {\boldsymbol a}_2$. 
Restricted to the triangular layer, the space group contains 
two translations $T_1$ along the ${\boldsymbol a}_1$ direction, 
$T_2$ along the ${\boldsymbol a}_2$ direction, a 
counterclockwise three-fold rotation $C_3$ around the lattice site, 
a two-fold rotation $C_2$ around ${{\boldsymbol a}_1 + {\boldsymbol a}_2}$, 
and the inversion $I$ at the lattice site.  
Their actions on the lattice indices are 
\begin{eqnarray}
    T_1&:& (x,y) \to (x+1, y), \\
    T_2&:& (x,y) \to (x, y+1), \\
    C_3&:& (x,y) \to (-y, x-y),\\
    C_2&:& (x,y) \to (y, x),\\
    I&:& (x, y) \to (-x, -y).
\end{eqnarray}

In the formulation introduced in the main text, 
we consider an equivalent set of generators, 
$\{ T_1, T_2, C_2, S_6 \}$, where the operation $S_6$ is {\sl defined} as
$S_6 \equiv C_3^{-1} I$ and acts on the lattice indices as
\begin{eqnarray}
    S_6: (x, y) \to (x-y, x).
\end{eqnarray}
It is evident that these two sets of generators are equivalent, since
we merely redefine the symmetry rather than introducing any new symmetry. 

The multiplication rule of this symmetry group is given in the main text.
For the convenience of the presentation below, we also list these rules here,
\begin{eqnarray}
T_1^{-1} T_2^{} T_1^{} T_2^{-1} &=&
T_1^{-1} T_2^{-1} T_1^{} T_2^{}  = 1 ,
\label{seq1}
\\
C_2^{-1} T_1^{} C_2^{} T_2^{-1} &=&
C_2^{-1} T_2^{} C_2^{} T_1^{-1}  = 1 ,
\label{seq2}
\\
S_6^{-1} T_1^{} C_6^{} T_2^{}   &=&
S_6^{-1} T_2^{} C_6^{} T_2^{-1} T_1^{-1} =1 ,
\label{seq3}
\\
(C_2^{})^2 = (C_6^{})^6 &=&
(S_6^{} C_2^{})^2 = 1.
\label{seq4}
\end{eqnarray}

Including the time reversal symmetry, we further have 
\begin{eqnarray}
T_1^{-1} \mathcal{T} T_1^{} \mathcal{T} & = &  
T_2^{-1} \mathcal{T} T_2^{} \mathcal{T} = 1 ,
\label{seq5}\\
C_2^{-1} \mathcal{T} C_2^{} \mathcal{T} & = &
S_6^{-1} \mathcal{T} S_6^{} \mathcal{T} = 1 ,
\label{seq6}\\
&& \mathcal{T}^2=1.
\label{seq7} 
\end{eqnarray}

\section{Projective symmetry group classification}
\label{ssec2}

As we describe in the main text, we consider the U(1) QSL. 
The spinon mean-field Hamiltonian has the following form 
\begin{eqnarray}
H_{\text{MF}}^{} = -
\sum_{ ({\boldsymbol r}{\boldsymbol r}')} \sum_{ \alpha\beta } \big[\,
t_{{\boldsymbol r}{\boldsymbol r}',\alpha\beta}^{} 
f^{\dagger}_{{\boldsymbol r} \alpha} f^{}_{\boldsymbol{r}'\beta} + h.c.
\,\big],
\label{seqmft}
\end{eqnarray}
where $t_{{\boldsymbol r}{\boldsymbol r}',\alpha\beta}$ is the 
spin-dependent hopping. With the extended Nambu spinor 
representation~\cite{Jason} 
$\Psi_{\boldsymbol{r}}^{} = (f^{}_{\boldsymbol{r}\uparrow}, 
f^{\dagger}_{\boldsymbol{r}\downarrow}, f^{}_{\boldsymbol{r}\downarrow}, 
- f^{\dagger}_{\boldsymbol{r}\uparrow})^T$, $H_{\text{MF}}^{}$
has a more compact form 
\begin{eqnarray}
H_{\text{MF}}^{} = - \frac{1}{2} 
\sum_{(\boldsymbol{r},\boldsymbol{r}')} 
\big[ 
\Psi_{\boldsymbol{r}}^{\dagger} 
u_{\boldsymbol{r} \boldsymbol{r}'}^{} 
\Psi^{}_{\boldsymbol{r}'}
+ h.c. \big],
\end{eqnarray}
where $u_{\boldsymbol{r} \boldsymbol{r}'}^{}$ is a hopping matrix  
that is related to $t_{{\boldsymbol r}{\boldsymbol r}',\alpha\beta}$,
\begin{eqnarray}
u_{\boldsymbol{r} \boldsymbol{r}'}^{} = \left(
\begin{array}{cccc}
t_{\boldsymbol{r} \boldsymbol{r}',\uparrow\uparrow} & 0 & 
t_{\boldsymbol{r} \boldsymbol{r}',\uparrow\downarrow} & 0 \\
0 & -t^{\ast}_{\boldsymbol{r} \boldsymbol{r}',\downarrow\downarrow} & 
0 &  t^{\ast}_{\boldsymbol{r} \boldsymbol{r}',\downarrow\uparrow} \\
t_{\boldsymbol{r} \boldsymbol{r}',\downarrow\uparrow} & 0 &
t_{\boldsymbol{r} \boldsymbol{r}',\downarrow\downarrow} & 0 \\
0 & t^{\ast}_{\boldsymbol{r} \boldsymbol{r}',\uparrow\downarrow} &
0 & - t^{\ast}_{\boldsymbol{r} \boldsymbol{r}',\uparrow\uparrow}
\end{array}
\right).
\end{eqnarray}

\begin{table}
\renewcommand\arraystretch{1.4}
\begin{tabular}{ccccc}
\hline\hline
U(1) QSL & $W^{T_1}_{\boldsymbol r}$ & $W^{T_2}_{\boldsymbol r}$
& $W^{C_2}_{\boldsymbol r}$ & $W^{S_6}_{\boldsymbol r}$
\\
\hline
U1A00 & $I_{2\times 2}^{}$ & $I_{2\times 2}^{}$ & $I_{2\times 2}^{}$ & $I_{2\times 2}^{}$
\\
U1A10 & $I_{2\times 2}^{}$ & $I_{2\times 2}^{}$ & $i\sigma^y$ & $I_{2\times 2}^{}$
\\
U1A01 & $I_{2\times 2}^{}$ & $I_{2\times 2}^{}$ & $I_{2\times 2}^{}$ & $i \sigma^y $
\\
U1A11 & $I_{2\times 2}^{}$ & $I_{2\times 2}^{}$ & $i\sigma^y$   &     $i \sigma^y $
\\
U1B00 & $I_{2\times 2}^{}$ & $(-1)^x I_{2\times 2}$ & $(-1)^{x y} I_{2\times 2}^{}$ & $(-1)^{x y
 - \frac{y(y-1)}{2}} I_{2\times 2}^{}$
\\
U1B10 & $I_{2\times 2}^{}$ & $(-1)^x I_{2\times 2}^{}$ & $i\sigma^y (-1)^{xy}$ & $(-1)^{x y -
\frac{ y(y-1)}{2}}I_{2\times 2}^{}$
\\
U1B01 & $I_{2\times 2}^{}$ & $(-1)^x I_{2\times 2}^{}$ & $(-1)^{x y}I_{2\times 2}^{}$ & $i\sigma^y(-1)^{x y -
\frac{ y(y-1)}{2}}$
\\
U1B11 & $I_{2\times 2}^{}$ & $(-1)^x I_{2\times 2}^{}$  & $i\sigma^y(-1)^{x y}$ & $i\sigma^y(-1)^{x y -
\frac{ y(y-1)}{2}}$
\\
\hline\hline
\end{tabular}
\caption{List of the gauge transformations for the
symmetry operations of the eight U(1) PSGs,
where $(x,y)$ is the coordinate in the oblique 
coordinate system. For time reversal symmetry, 
all PSGs have the same gauge transformation 
$W_{\boldsymbol{r}}^{\mathcal{T}} = I_{2\times 2}$. }
\label{tab}
\end{table}

\subsection{Spatial symmetry}

First of all, the gauge transformation and spin rotation are commutative. 
So in the PSG classification, we only need to focus on the gauge part of 
the PSG transformation. In the canonical gauge $\text{IGG}=\{ I_{2\times 2} 
\otimes e^{i \phi \sigma^z}| \phi \in [0,2\pi) \}$, the gauge transformation 
associated with a given symmetry operation $\mathcal{O}$ takes the form 
\begin{eqnarray}
\mathcal{G}^{\mathcal O}_{\boldsymbol{r}}  =
I_{2\times 2} \otimes W^{\mathcal O}_{\boldsymbol{r}}
\equiv I_{2\times 2} \otimes \big[ (i \sigma^x)^{n^{}_{\mathcal O}}
e^{i \phi^{}_{\mathcal O} [{\boldsymbol r}] \sigma^z} \big],
\end{eqnarray}
where ${n_{\mathcal O}^{} = 0,1}$.  For the symmetry multiplication rule 
${\mathcal{O}_1 \mathcal{O}_2 \mathcal{O}_3 \mathcal{O}_4 =1}$ where 
$\mathcal{O}_i$ is an unitary transformation, the corresponding
PSG relation becomes 
\begin{equation}
{\mathcal G}^{\mathcal{O}_1}_{\boldsymbol{r}}
{\mathcal G}^{\mathcal{O}_2}_{\mathcal{O}_2\mathcal{O}_3\mathcal{O}_4(\boldsymbol{r})}
{\mathcal G}^{\mathcal{O}_3}_{\mathcal{O}_3\mathcal{O}_4(\boldsymbol{r})}
{\mathcal G}^{\mathcal{O}_4}_{\mathcal{O}_4(\boldsymbol{r})} \in \text{IGG}
\end{equation}
or equivalently, 
\begin{eqnarray}
&& {W}^{\mathcal{O}_1}_{\boldsymbol{r}}
{W}^{\mathcal{O}_2}_{\mathcal{O}_2\mathcal{O}_3\mathcal{O}_4(\boldsymbol{r})}
{W}^{\mathcal{O}_3}_{\mathcal{O}_3\mathcal{O}_4(\boldsymbol{r})}
{W}^{\mathcal{O}_4}_{\mathcal{O}_4(\boldsymbol{r})}
\nonumber 
\\
&& \quad\quad\quad\quad  \in  \{ e^{i \phi \sigma^z} | \phi \in [0,2\pi ) \}.
\end{eqnarray}

We start with $T_1$ and $T_2$, where 
\begin{eqnarray}
W^{T_1}_{\boldsymbol r} &=& (i\sigma^x)^{n_{T_1}^{}} ,
\\
W^{T_2}_{\boldsymbol r} &=& (i\sigma^x)^{n_{T_2}^{}} 
e^{i \phi_{T_2}^{} [ {\boldsymbol r} ] \sigma^z}. 
\end{eqnarray}
Through Eq.~(\ref{seq2}) that connects $T_1$ and $T_2$, 
one immediately has ${n_{T_1} = n_{T_2}}$. From 
Eq.~(\ref{seq3}) where the total number of $T_1$ and $T_2$ 
is odd, one immediately has ${n_{T_1} = n_{T_2} = 0}$. 
So we have 
\begin{eqnarray}
W^{T_1}_{\boldsymbol r} &=& 1, \\ 
W^{T_2}_{\boldsymbol r} &=& e^{i \phi_{T_2} [x,y]\sigma^z}. 
\end{eqnarray}
Using Eq.~(\ref{seq1}), we have 
\begin{eqnarray}
[W^{T_1} T_1]^{-1} 
[W^{T_2} T_2] 
[W^{T_1} T_1]  
[W^{T_2} T_2]^{-1}
\nonumber 
\\
= T_1^{-1} (W^{T_1})^{-1} W^{T_2} T_2 W^{T_1} T_1
T_2^{-1} W_{T_2}^{-1} 
\nonumber 
\\ 
\in \{
 e^{i \phi \sigma^z}| \phi \in [0,2\pi) \},
\end{eqnarray}
which leads to the result 
\begin{eqnarray}
\phi_{T_2}^{} [x+1,y] - \phi_{T_2}^{} [x,y] \equiv \phi_1
\end{eqnarray}
with $\phi_1$ to be determined. Since it is always possible to choose a gauge
such that $\phi_{T_2} [0,y]=0$, then we have $\phi_{T_2} [x,y]= \phi_1 x$. 

Similarly, $T_1^{-1}T_2^{-1} T_1 T_2 =1$ leads to 
\begin{eqnarray}
\phi_{T_2}^{} [x+1,y+1] 
-\phi_{T_2}^{} [x,y+1] = \phi_2. 
\end{eqnarray}
It is ready to find $\phi_2=\phi_1$. 

We continue to find $W^{S_6}_{\boldsymbol{r}}$ and 
$W^{C_2}_{\boldsymbol{r}}$. For the operation $S_6$ with 
$W^{S_6}_{\boldsymbol{r}} = (i \sigma^x)^{n_{S_6}} e^{i \phi_{S_6} [x,y] \sigma^z}$,
Eq.~(\ref{seq3}) leads to 
\begin{eqnarray}
- \phi_{S_6} [T_1(\boldsymbol{r})] + \phi_{S_6} [{\boldsymbol{r}}] &=& -\phi_1 y + \phi_3 ,\\
- \phi_{S_6} [T_2(\boldsymbol{r})] + \phi_{S_6} [{\boldsymbol{r}}] &=& \phi_4 - \phi_1 x+\phi_1 y ,
\end{eqnarray}
for $n_{S_6} =0$, and 
\begin{eqnarray}
- \phi_{S_6} [T_1(\boldsymbol{r})] + \phi_{S_6} [{\boldsymbol{r}}] &=& -\phi_1 y + \phi_3
\\
- \phi_{S_6} [T_2(\boldsymbol{r})] + \phi_{S_6} [{\boldsymbol{r}}] &=& \phi_4 
+ \phi_1 x + \phi_1 y.
\end{eqnarray}
for $n_{S_6} = 1$. So we obtain 
\begin{eqnarray}
&& {\text{when  }}n_{S_6} = 0,  \nonumber \\
&&  \phi_{S_6}[{\boldsymbol r}] = \phi_1 x y- \phi_3 x-\phi_4 y -\frac{\phi_1 y(y-1)}{2}
\\
&& {\text{when  }}n_{S_6} = 1,   \nonumber \\  
&& \phi_{S_6}[{\boldsymbol r}] = \phi_1 x y- \phi_3 x-\phi_4 y -\frac{\phi_1 y(y-1)}{2}.
\end{eqnarray}
For $n_{S_6}=1$, we further require $\phi_1 = 0,\pi$. $S_6^6 = 1$ is 
automatically satisfied with the above relations for both $n_{S_6}=0$
and $n_{S_6} =1$. 

For $W^{C_2}_{\boldsymbol{r}}$ with
$W^{C_2}_{\boldsymbol{r}} = (i \sigma^x)^{n_{C_2}} e^{i \phi_{C_2} [x,y] \sigma^z}$, 
we need to consider two separate cases with $n_{c_2} = 0, 1$, respectively. 
If $n_{C_2} =0$, Eq.~(\ref{seq2}) leads to 
\begin{eqnarray}
-\phi_{T_2} [C_2^{-1} T_1 ({\boldsymbol r})] - \phi_{C_2} [T_1  ({\boldsymbol r})] 
+ \phi_{C_2} [{\boldsymbol r}] &=& \phi_5 , \\
-\phi_{C_2} [ T_2 ({\boldsymbol r})] + \phi_{T_2} [T_2 ({\boldsymbol r})] 
+ \phi_{C_2} [{\boldsymbol r}] &=& \phi_6. 
\end{eqnarray}
So we obtain $\phi_{C_2} [x,y] = -\phi_5 x - \phi_6 y - x y \phi_1$ and 
$\phi_1 = 0,\pi$ for $n_{C_2} =0$.   
Similary, for $n_{C_2} = 1$, we obtain 
$\phi_{C_2} [x,y] = -\phi_5 x - \phi_6 y - x y \phi_1$. 

Using $C_2^2 =1$, we further have
$\phi_6 =- \phi_5$ for $n_{C_2} = 0$, 
and $\phi_6 = \phi_6$ for $n_{C_2} =1$.  
So we arrive at the result
\begin{eqnarray}
n_{C_2} &=& 0, \quad \phi_{C_2}[x,y] = -\phi_5(x-y) - xy \phi_1, \\
n_{C_2} &=& 1, \quad \phi_{C_2}[x,y] = -\phi_5(x+y) - xy \phi_1 .
\end{eqnarray}
Here, to simplify the above expression, we choose a pure gauge tranformation
$\tilde{W}^a_{\boldsymbol r} = e^{ix \sigma^z \phi_5}$. Under the pure gauge
transformation, the gauge part of the PSG transforms as 
\begin{eqnarray}
W^{\mathcal O}_{\boldsymbol r} \rightarrow \tilde{W}^a_{\boldsymbol r} 
W^{\mathcal O}_{\boldsymbol r} \tilde{W}^{a\dagger}_{{\mathcal O}^{-1}(\boldsymbol r)}.
\end{eqnarray}
Clearly $\tilde{W}^a_{\boldsymbol r}$ only modifies $W^{T_1}$ and $W^{T_2}$ by an overall
phase shift, but ${W}^{C_2}_{\boldsymbol r}$ becomes
\begin{eqnarray}
{W}^{C_2}_{\boldsymbol r} = (i \sigma^x)^{n_{C_2}} e^{-i x y \phi_1 \sigma^z}
\end{eqnarray}
for both $n_{C_2} = 0, 1$, except that we require $\phi_1 =0,\pi$ for $n_{C_2} = 0$.

For the relation $(S_6 C_2)^2 =1$, we need to consider 
the four cases with $n_{S_6} = 0,1$ and $n_{C_2} = 0,1$. 

For $n_{S_6} = n_{C_2} = 0$, we have $\phi_1 = \pi$, 
and $(S_6 C_2)^2 =1$ gives $\phi_3 + 2\phi_4 =0$. 
We then introduce a pure gauge transformation 
$\tilde{W}^b_{\boldsymbol r}$, 
\begin{eqnarray}
\tilde{W}^b_{\boldsymbol r} = e^{-i (x+y)\phi_4 \sigma^z}.
\end{eqnarray}
After applying $\tilde{W}^b_{\boldsymbol r} $, we have 
\begin{eqnarray}
\phi_{C_2} &=& - xy \phi_1,
\\
\phi_{S_6} &=& x y\phi_1- \phi_1 \frac{y(y-1)}{2}
\end{eqnarray}
with $\phi_1 =0, \pi$. 

For $n_{S_6} =0$ and $n_{C_2} = 1$, we obtain $\phi_3 =0$. 
We introduce a pure gauge transformation
$\tilde{W}^c_{\boldsymbol r} $,
\begin{eqnarray}
\tilde{W}^c_{\boldsymbol r} = e^{-i (x-y) \phi_4 \sigma^z}. 
\end{eqnarray}
After applying $\tilde{W}^b_{\boldsymbol r} $, we have 
\begin{eqnarray}
\phi_{C_2} &=& - xy \phi_1,
\\
\phi_{S_6} &=& x y\phi_1- \phi_1 \frac{y(y-1)}{2}. 
\end{eqnarray}

For $n_{S_6} =1$ and $n_{C_2} = 0$, we obtain 
$ \phi_3 =0$. We apply a pure gauge transformation
$\tilde{W}^b_{\boldsymbol r} $ and obtain
\begin{eqnarray}
\phi_{C_2} &=& - x y \phi_1,
\\
\phi_{S_6} &=& x y\phi_1 - \phi_1 \frac{y(y-1)}{2}. 
\end{eqnarray}

For $n_{S_6} =1$ and $n_{C_2} = 1$, 
we obtain $\phi_3 + 2\phi_4 = 0$. 
We apply a pure gauge transformation
$\tilde{W}^c_{\boldsymbol r}$ and obtain
\begin{eqnarray}
\phi_{C_2} &=& - x y \phi_1,
\\
\phi_{S_6} &=& x y\phi_1- \phi_1 \frac{y(y-1)}{2}. 
\end{eqnarray}

In summary, we have 
\begin{eqnarray}
W^{T_1}_{\boldsymbol r} =1 ,\quad W^{T_2}_{\boldsymbol r} = e^{i \phi_1 x}.
\end{eqnarray}
and 
\begin{eqnarray}
W^{C_2}_{\boldsymbol r} &=& (i\sigma^x)^{n_{C_2}^{}}  
e^{-i \phi_1 x y \sigma^z},
\\
W^{S_6}_{\boldsymbol r} &=& (i\sigma^x)^{n_{S_6}^{}}  
e^{i\phi_1 [ x y - \frac{y(y-1)}{2} ] \sigma^z },
\end{eqnarray}
where ${\phi_1 = 0 ,\pi}$ for $n_{C_2} = 0$ or $n_{S_6} = 1$. 

\subsection{Time reversal symmetry}

Because time reversal is an antiunitary symmetry, 
the product ${\mathcal O}^{-1} {\mathcal T}^{-1} {\mathcal O} {\mathcal T}$ becomes  
\begin{eqnarray}
(W^{\mathcal O}_{\boldsymbol r})^{\dagger} 
[(W^{\mathcal T}_{\boldsymbol r})^{\dagger} W^{\mathcal O}_{\boldsymbol r} W^{\mathcal T}_{\mathcal{O}^{-1} ({\boldsymbol r})}]^{\ast}   
\end{eqnarray}
for the PSGs, where $W^{\mathcal T}$ is the gauge transformation associated with the time reversal. 
We here redefine 
\begin{eqnarray}
W^{\mathcal T}_{\boldsymbol r} = \bar{W}^{\mathcal T}_{\boldsymbol r} (i \sigma^y),
\end{eqnarray}
so that 
\begin{eqnarray}
{\mathcal O}^{-1} {\mathcal T}^{-1} {\mathcal O} {\mathcal T} \rightarrow 
(W^{\mathcal O}_{\boldsymbol r})^{\dagger} 
(\bar{W}^{\mathcal T}_{\boldsymbol r} )^{\dagger}
W^{\mathcal O}_{\boldsymbol r}
\bar{W}^{\mathcal T}_{\mathcal{O}^{-1}({\boldsymbol r})} . 
\end{eqnarray}
$\bar{W}^{\mathcal T}_{\boldsymbol r}$ has the general form 
$\bar{W}^{\mathcal T}_{\boldsymbol r} 
= (i\sigma^x)^{n_{\mathcal T}} 
e^{i \phi_{\mathcal T}[{\boldsymbol r}] \sigma^z}$.   

We start with $n_{\mathcal T} = 0$. 
The relation in Eq.~(\ref{seq5}) leads to 
\begin{eqnarray}
\phi_{\mathcal T} [x,y] - \phi_{\mathcal T}[x-1,y] & = & - \phi_7 ,\\
\phi_{\mathcal T} [x,y+1] - \phi_{\mathcal T}[x,y] & = & - \phi_8 ,
\end{eqnarray}
so we have $\phi_{\mathcal T} [x,y] = - \phi_7 x - \phi_8 y$. 
Applying this result to Eq.~(\ref{seq6}), we have 
\begin{eqnarray}
-\phi_{C_2} [ y,x] - \phi_{\mathcal T}[y,x] + \phi_{C_2}[y,x]&& \nonumber \\
+ \phi_{\mathcal T}[x,y]  &=& \phi_{9},
\nonumber \\
-\phi_{S_6}[x,y]-\phi_{\mathcal T}[x,y] + \phi_{S_6}[x,y] && \nonumber \\
 + \phi_{\mathcal T}[y,-x+y] &=& \phi_{10}, 
\end{eqnarray}
for ${n_{C_2} = n_{S_6} = 0}$. 
The above equations give ${\phi_7=\phi_8=0}$, 
so we have ${\bar{W}^{\mathcal T}_{\boldsymbol r} = 1}$. 
Other cases can be obtained likewise. 
We find that for both ${n_{\mathcal T} =0}$ 
and ${n_{\mathcal T} = 1}$, there is ${\phi_{\mathcal T} [x,y]=0}$ 
and ${\phi_1 = 0,\pi}$. So we have 
\begin{eqnarray}
\bar{W}^{\mathcal T}_{\boldsymbol r} = 1, i\sigma^y,
\end{eqnarray}
where we have used a global and uniform rotation 
$e^{i \frac{\pi}{4} \sigma^z}$ to rotate $\sigma^x$ to 
the basis of $\sigma^y$. 

Including the time reversal, there are 16 PSG solutions. 
But for $\bar{W}^{\mathcal T}_{\boldsymbol r} = 1$, the 
mean-field ansatz is found to vanish everythere. This makes sense
as these PSGs have ${\mathcal T}^2 =1 $ for the fermionic spinons
that are expected to Kramers doublets. So only 8 of them with 
${\mathcal T}^2 =-1$ for the spinons survive. Replacing 
$e^{i \phi_1 \sigma^z}$ with $\pm 1$, we present the PSG 
solutions in the table of the main text.

\section{Spinon band structures and mean-field Hamiltonians}
\label{ssec3}

As we establish in the previous section and the main text, 
there are four U1A PSGs and four U1B PSGs. In the main text, 
we have argued that the experimental resuls in YbMgGaO$_4$ 
is against the U1B states. So here we focus on the U1A states. 
From the U1A PSGs, it is straight to obtain the spinon transformations. 
We list the results in Tab.~\ref{stab1}.

\begin{table*}
\caption{The transformation for the spinons under four U1A PSGs that are labeled by
U1A$n_{C_2}n_{S_6}$. }
\begin{tabular}{ccccc}
\hline\hline
U(1) PSGs & $T_1$ & $T_2$ & {$C_2$} & $S_6$  \\
\hline
U1A00 & 
$\begin{array}{cc}
f_{(x,y),\uparrow} \rightarrow f_{(x+1,y),\uparrow}  
\\ 
f_{(x,y),\downarrow} \rightarrow f_{(x+1,y),\downarrow}
\end{array}$ & 
$\begin{array}{cc}
f_{(x,y),\uparrow} \rightarrow f_{(x,y+1),\uparrow}  
\\ 
f_{(x,y),\downarrow} \rightarrow f_{(x,y+1),\downarrow}
\end{array}$ 
 & 
$\begin{array}{cc}
f_{(x,y),\uparrow} \rightarrow e^{i\frac{\pi}{6}} f_{(y,x),\downarrow}  
\\ 
f_{(x,y),\downarrow} \rightarrow e^{i\frac{5\pi}{6}} f_{(y,x),\uparrow}
\end{array}$  
 & 
 $\begin{array}{cc}
f_{(x,y),\uparrow} \rightarrow e^{-i\frac{\pi}{3}} f_{(x-y,x),\uparrow}  
\\ 
f_{(x,y),\downarrow} \rightarrow e^{+i\frac{\pi}{3}} f_{(x-y,x),\downarrow}
\end{array}$   
\\
\hline
U1A10 & 
$\begin{array}{cc}
f_{(x,y),\uparrow} \rightarrow f_{(x+1,y),\uparrow}  
\\ 
f_{(x,y),\downarrow} \rightarrow f_{(x+1,y),\downarrow}
\end{array}$ & 
$\begin{array}{cc}
f_{(x,y),\uparrow} \rightarrow f_{(x,y+1),\uparrow}  
\\ 
f_{(x,y),\downarrow} \rightarrow f_{(x,y+1),\downarrow}
\end{array}$ 
 & 
$\begin{array}{cc}
f_{(x,y),\uparrow} \rightarrow e^{i\frac{\pi}{6}} f^{\dagger}_{(y,x),\uparrow}  
\\ 
f_{(x,y),\downarrow} \rightarrow e^{-i\frac{\pi}{6}} f^{\dagger}_{(y,x),\downarrow}
\end{array}$
 & 
 $\begin{array}{cc}
f_{(x,y),\uparrow} \rightarrow e^{-i\frac{\pi}{3}} f_{(x-y,x),\uparrow}  
\\ 
f_{(x,y),\downarrow} \rightarrow e^{+i\frac{\pi}{3}} f_{(x-y,x),\downarrow}
\end{array}$    
\\
\hline
U1A01 & 
$\begin{array}{cc}
f_{(x,y),\uparrow} \rightarrow f_{(x+1,y),\uparrow}  
\\ 
f_{(x,y),\downarrow} \rightarrow f_{(x+1,y),\downarrow}
\end{array}$ & 
$\begin{array}{cc}
f_{(x,y),\uparrow} \rightarrow f_{(x,y+1),\uparrow}  
\\ 
f_{(x,y),\downarrow} \rightarrow f_{(x,y+1),\downarrow}
\end{array}$ 
 & 
$\begin{array}{cc}
f_{(x,y),\uparrow} \rightarrow e^{i\frac{\pi}{6}} f_{(y,x),\downarrow}  
\\ 
f_{(x,y),\downarrow} \rightarrow e^{i\frac{5\pi}{6}} f_{(y,x),\uparrow}
\end{array}$  
 & 
 $\begin{array}{cc}
f_{(x,y),\uparrow} \rightarrow -e^{-i\frac{\pi}{3}} f^{\dagger}_{(x-y,x),\downarrow}  
\\ 
f_{(x,y),\downarrow} \rightarrow e^{+i\frac{\pi}{3}} f^{\dagger}_{(x-y,x),\uparrow}
\end{array}$   
\\
\hline
U1A11 & 
$\begin{array}{cc}
f_{(x,y),\uparrow} \rightarrow f_{(x+1,y),\uparrow}  
\\ 
f_{(x,y),\downarrow} \rightarrow f_{(x+1,y),\downarrow}
\end{array}$ & 
$\begin{array}{cc}
f_{(x,y),\uparrow} \rightarrow f_{(x,y+1),\uparrow}  
\\ 
f_{(x,y),\downarrow} \rightarrow f_{(x,y+1),\downarrow}
\end{array}$ 
 & 
$\begin{array}{cc}
f_{(x,y),\uparrow} \rightarrow e^{i\frac{\pi}{6}} f^{\dagger}_{(y,x),\uparrow}  
\\ 
f_{(x,y),\downarrow} \rightarrow e^{-i\frac{\pi}{6}} f^{\dagger}_{(y,x),\downarrow}
\end{array}$  
 & 
 $\begin{array}{cc}
f_{(x,y),\uparrow} \rightarrow -e^{-i\frac{\pi}{3}} f^{\dagger}_{(x-y,x),\downarrow}  
\\ 
f_{(x,y),\downarrow} \rightarrow e^{+i\frac{\pi}{3}} f^{\dagger}_{(x-y,x),\uparrow}
\end{array}$ \\
\hline\hline
\end{tabular}
\label{stab1}
\end{table*}

\subsection{Spinon band structures}

Using Tab.~\ref{stab1}, we obtain the spinon mean-field Hamiltonian. 
In particular, the U1A10 state gives vanishing spinon hoppings 
on the first and second neighbors, and the U1A01 state gives an 
isotropic spinon hopping on both first and second neighbors.  
The U1A10 state, as we described in the main text, 
has symmetry protected band touchings at the $\Gamma$, M and K points.
The U1A11 state has symmetry protected band touchings at the $\Gamma$
and M points. 

For the U1A10 state, the spinon mean-field Hamiltonian has the form 
\begin{equation}
H_{\text{MF}}^{\text{U1A01}} = \sum_{\boldsymbol{k}}
h_{\alpha\beta}^{} ({\boldsymbol k})
f^{\dagger}_{{\boldsymbol k}\alpha}
f^{}_{{\boldsymbol k}\beta}, 
\end{equation}
where $h_{\alpha\beta}^{} ({\boldsymbol k})$ is given by 
\begin{eqnarray}
h ({\boldsymbol k})
= d_0 ({\boldsymbol k}) I_{2 \times 2}^{}
+ \sum_{\mu=1}^3 {d}_{\mu} ({\boldsymbol k})
{{\sigma}^{\mu}} .
\end{eqnarray}
In the main text, we have used $(S_6)^3$ and $\mathcal{T}$
to show $d_0({\boldsymbol k}) = 0$ and the band touchings 
at $\Gamma$ and M. To account for the band touching at the 
K point, we need to use $S_6$ and $C_2$. Under $S_6$, 
\begin{eqnarray}
    S_6 \CH S_6^{-1}
    &=& \sum_{\boldsymbol k} \big[ e^{\frac{i2\pi}{3}} 
    h(-S_6^{-1}({\boldsymbol k}))_{\ua\da}
    f^\dg_{{\boldsymbol k}\ua}f^\pg_{{\boldsymbol k}\da} + h.c.\big] \nn
    &=& \CH,
\end{eqnarray}
where ${h({\boldsymbol k})_{\ua\da} = d_x({\boldsymbol k}) - i d_y({\boldsymbol k})}$.
Since K is invariant under $S_6$,
\begin{eqnarray}
    d_x({\rm K}) - i d_y({\rm K}) = 
    e^{\frac{i2\pi}{3}} [ d_x({\rm K}) - i d_y({\rm K}) ] ,
\end{eqnarray}
hence $d_x({\rm K}) = d_y({\rm K}) = 0$.

The $C_2$ symmetry constraints the $d_z$ term, we have 
\begin{eqnarray}
C_2^\pg \CH C_2^{-1}
    &=& \sum_{\boldsymbol k} d_z(C_2^{-1}({\boldsymbol k})) f^\dg_{{\boldsymbol k}\da} f^\pg_{{\boldsymbol k}\da} - d_z(C_2^{-1}({\boldsymbol k})) f^\dg_{{\boldsymbol k}\ua} f^\pg_{{\boldsymbol k} \ua} \nn
    &=& \CH.
\end{eqnarray}
Since K is also invariant under $C_2$, we obtain 
$   d_z({\rm K}) = -d_z({\rm K})$.
Hence $d_z({\rm K}) = 0$. We conclude 
that $h({\rm K}) = 0$ and there exists 
a band touching at K.
  
For the U1A11 state, $\mathcal{T}$ and $S_6$ are implemented 
in the same way as the U1A01 state, and we arrive at 
the same conclusion that there are band touchings at 
the $\Gamma$ and M points. At the K point, however, 
the band structure is generally gapped due to a nonzero $d_z$.

\subsection{Spinon mean-field Hamiltonians}

The U1A00 state has the isotropic spinon hoppings on first and second 
neighboring bonds, and the mean-field Hamiltonain $H^{\text{U1A00}}_{\text{MF}}$
has already been given in the main text. This states gives a large 
spinon Fermi surface in the Brioullin zone.  
The spinon mean-field states of the U1A01 state and the U1A11 state are given by
\begin{widetext}
\begin{eqnarray}
    H_{\rm MF}^{\rm U1A01} &=& \sum_{x, y} t_1 \Big[
    -i f^\dg_{(x+1,y),\ua}f^\pg_{(x,y),\da}
    -i f^\dg_{(x+1,y),\da}f^\pg_{(x,y),\ua}
    -e^{-\frac{i\pi}{6}} f^\dg_{(x,y+1),\ua}f^\pg_{(x,y),\da} \nn
    &&\quad \quad \quad \quad
    +e^{\frac{i\pi}{6}} f^\dg_{(x,y+1),\da}f^\pg_{(x,y),\ua}
    -e^{\frac{i\pi}{6}} f^\dg_{(x+1,y+1),\ua}f^\pg_{(x,y),\da}
    +e^{-\frac{i\pi}{6}} f^\dg_{(x+1,y+1),\da}f^\pg_{(x,y),\ua}
    +h.c.
    \Big]\nn
    && \quad \quad + t_2 \Big[
      e^{\frac{i2\pi}{3}} f^\dg_{(x+1,y-1),\ua}f^\pg_{(x,y),\da}
    + e^{\frac{i\pi}{3}}  f^\dg_{(x+1,y-1),\da}f^\pg_{(x,y),\ua}
    + f^\dg_{(x+1,y+2),\ua}f^\pg_{(x,y),\da}\nn
    &&\quad \quad \quad \quad
    - f^\dg_{(x+1,y+2),\da}f^\pg_{(x,y),\ua}
    + e^{\frac{i\pi}{3}} f^\dg_{(x+2,y+1),\ua}f^\pg_{(x,y),\da}
    + e^{\frac{i2\pi}{3}} f^\dg_{(x+2,y+1),\da}f^\pg_{(x,y),\ua}
    + h.c.
    \Big],
\end{eqnarray} 
and   
    \begin{eqnarray}
    H_{\rm MF}^{\rm U1A11} &=& \sum_{x, y} t_1 \Big[
     i f^\dg_{(x+1,y),\ua}f^\pg_{(x,y),\ua}
    -i f^\dg_{(x+1,y),\da}f^\pg_{(x,y),\da}
    +i f^\dg_{(x,y+1),\ua}f^\pg_{(x,y),\ua}\nn
    &&\quad \quad \quad \quad
    -i f^\dg_{(x,y+1),\da}f^\pg_{(x,y),\da}
    -i f^\dg_{(x+1,y+1),\ua}f^\pg_{(x,y),\ua}
    +i f^\dg_{(x+1,y+1),\da}f^\pg_{(x,y),\da}
    +h.c.
    \Big]\nn
    && \quad \quad + t_1^\p \Big[
    - f^\dg_{(x+1,y),\ua}f^\pg_{(x,y),\da}
    + f^\dg_{(x+1,y),\da}f^\pg_{(x,y),\ua}
    + e^{\frac{i\pi}{3}} f^\dg_{(x,y+1),\ua}f^\pg_{(x,y),\da}\nn
    &&\quad \quad \quad \quad
    + e^{\frac{i2\pi}{3}} f^\dg_{(x,y+1),\da}f^\pg_{(x,y),\ua}
    + e^{\frac{i2\pi}{3}} f^\dg_{(x+1,y+1),\ua}f^\pg_{(x,y),\da}
    + e^{\frac{i\pi}{3}} f^\dg_{(x+1,y+1),\da}f^\pg_{(x,y),\ua}
    + h.c.
    \Big]\nn
    && \quad \quad + t_2 \Big[
      e^{\frac{i\pi}{6}} f^\dg_{(x+1,y-1),\ua}f^\pg_{(x,y),\da}
    + e^{\frac{i5\pi}{6}}  f^\dg_{(x+1,y-1),\da}f^\pg_{(x,y),\ua}
    -i f^\dg_{(x+1,y+2),\ua}f^\pg_{(x,y),\da}\nn
    &&\quad \quad \quad \quad
    -i f^\dg_{(x+1,y+2),\da}f^\pg_{(x,y),\ua}
    +e^{\frac{i5\pi}{6}}  f^\dg_{(x-2,y-1),\ua}f^\pg_{(x,y),\da}
    + e^{\frac{i\pi}{6}} f^\dg_{(x-2,y-1),\da}f^\pg_{(x,y),\ua}
    + h.c.
    \Big],
\end{eqnarray}
\end{widetext}
where in both Hamiltonians $t_1$, $t_1^\p$ denote the 
first neighbor hoppings and $t_2$ denotes the second neighbor hopping.

The band structures for specific choices of the hopping parameters are plotted in
the main text. Clearly, we observe the band touchings at the $\Gamma$, M and K points
for the U1A01 state, and band touchings at the $\Gamma$ and M points for the U1A11
state.

\section{The U1A00 state and the spectroscopic results}

\subsection{Free spinon mean-field theory}

\begin{figure*}[thp]
\centering
\includegraphics[width=.24\textwidth]{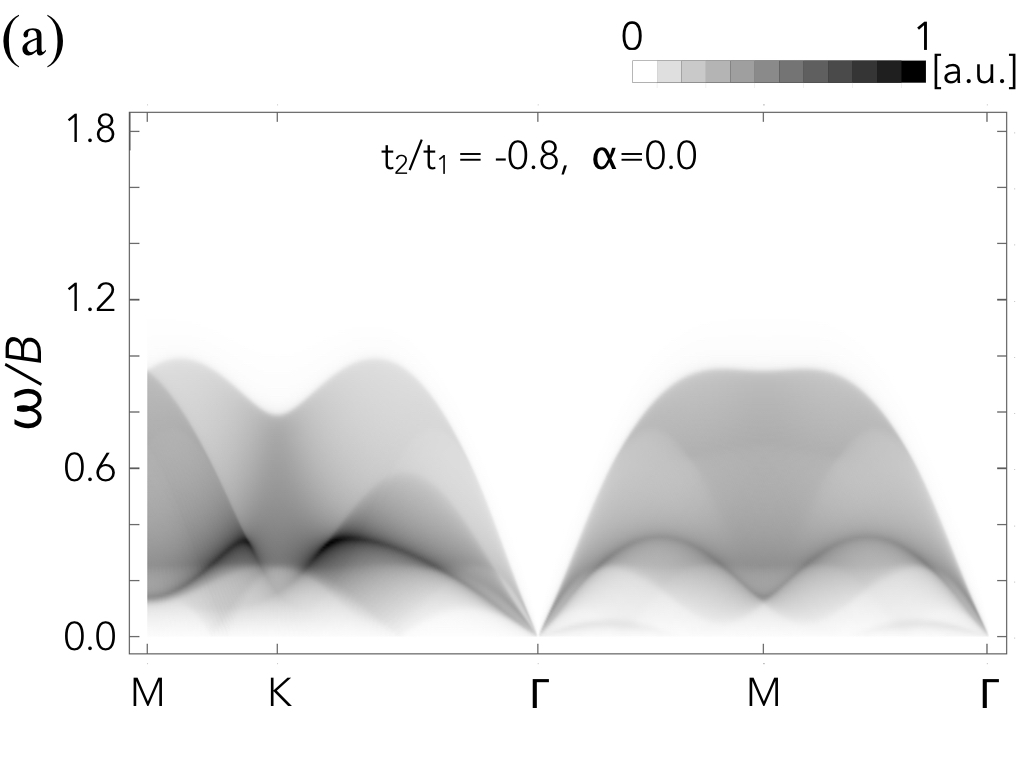}
\includegraphics[width=.24\textwidth]{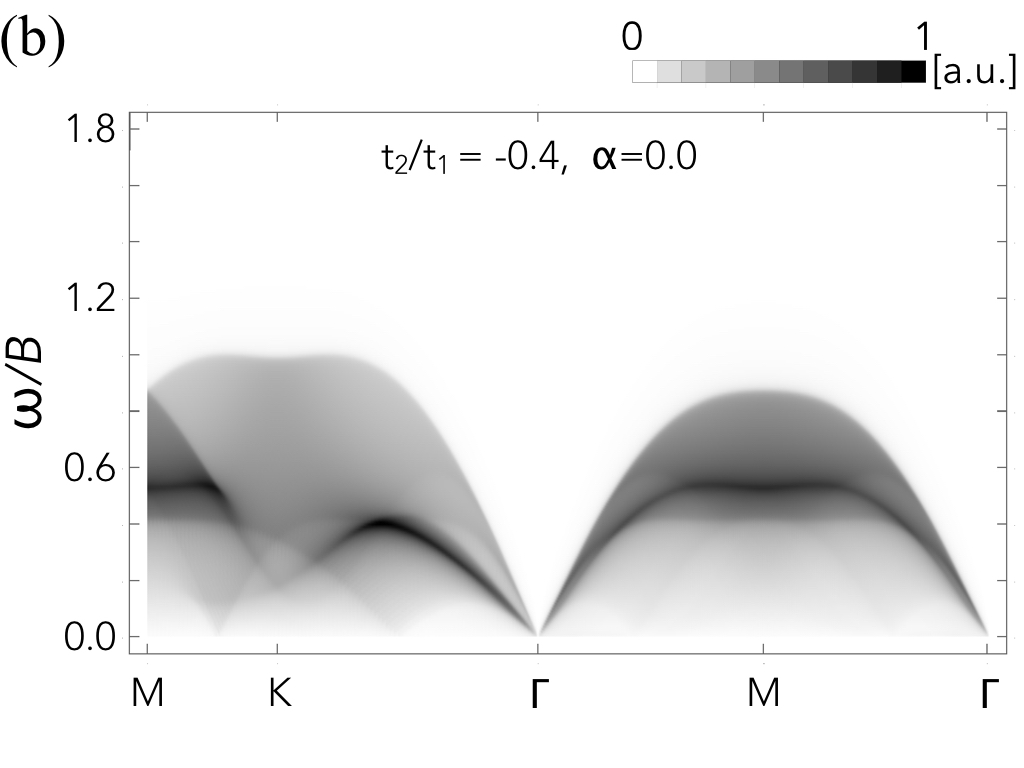}
\includegraphics[width=.24\textwidth]{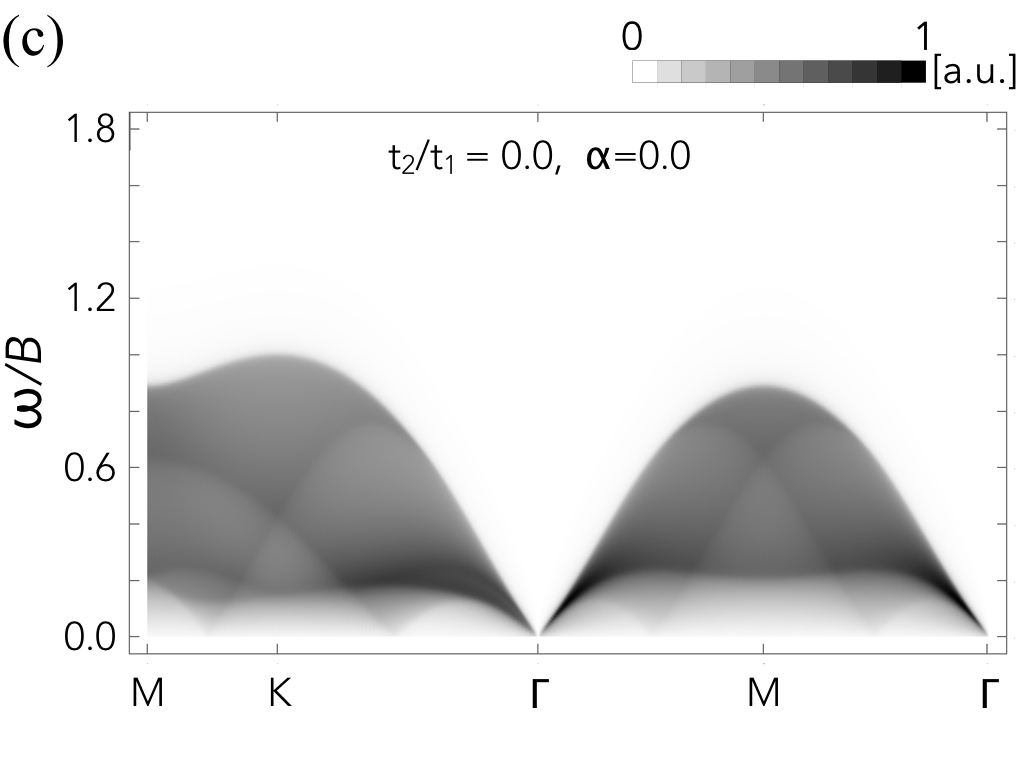}
\includegraphics[width=.24\textwidth]{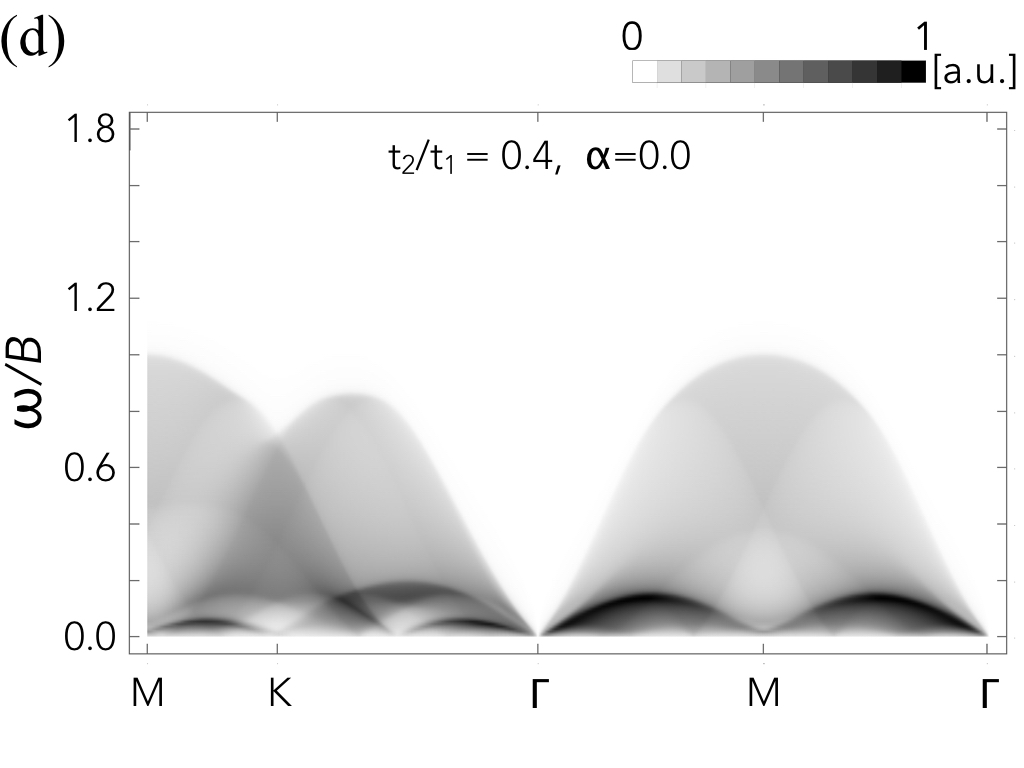}
\includegraphics[width=.24\textwidth]{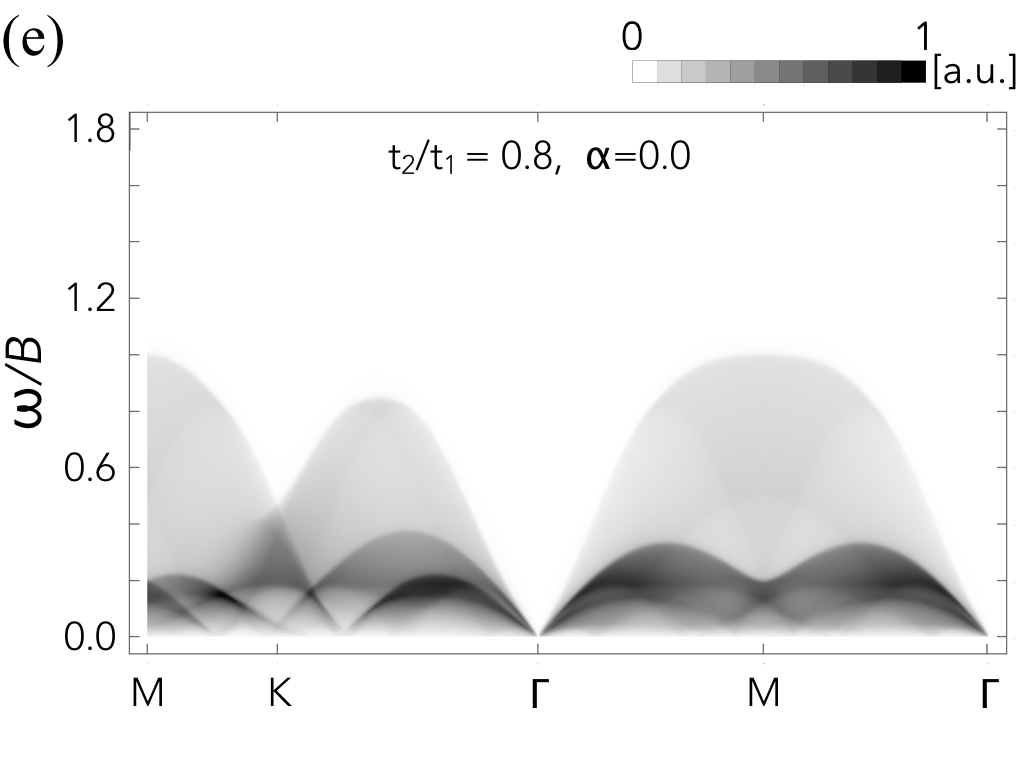}
\includegraphics[width=.24\textwidth]{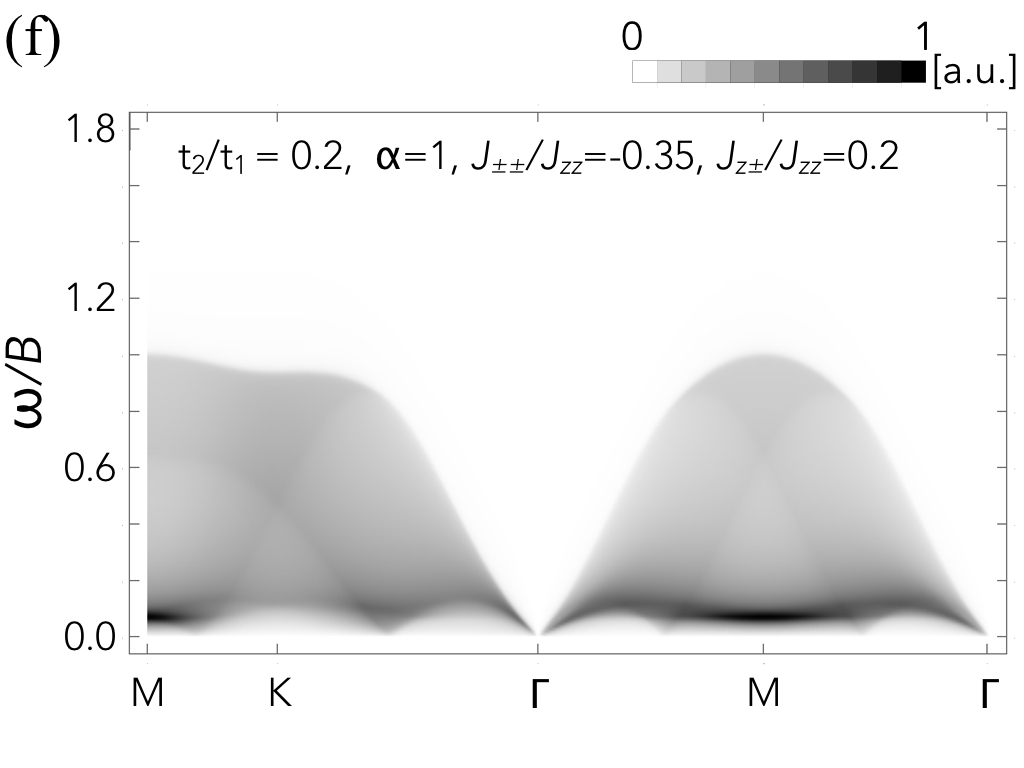}
\includegraphics[width=.24\textwidth]{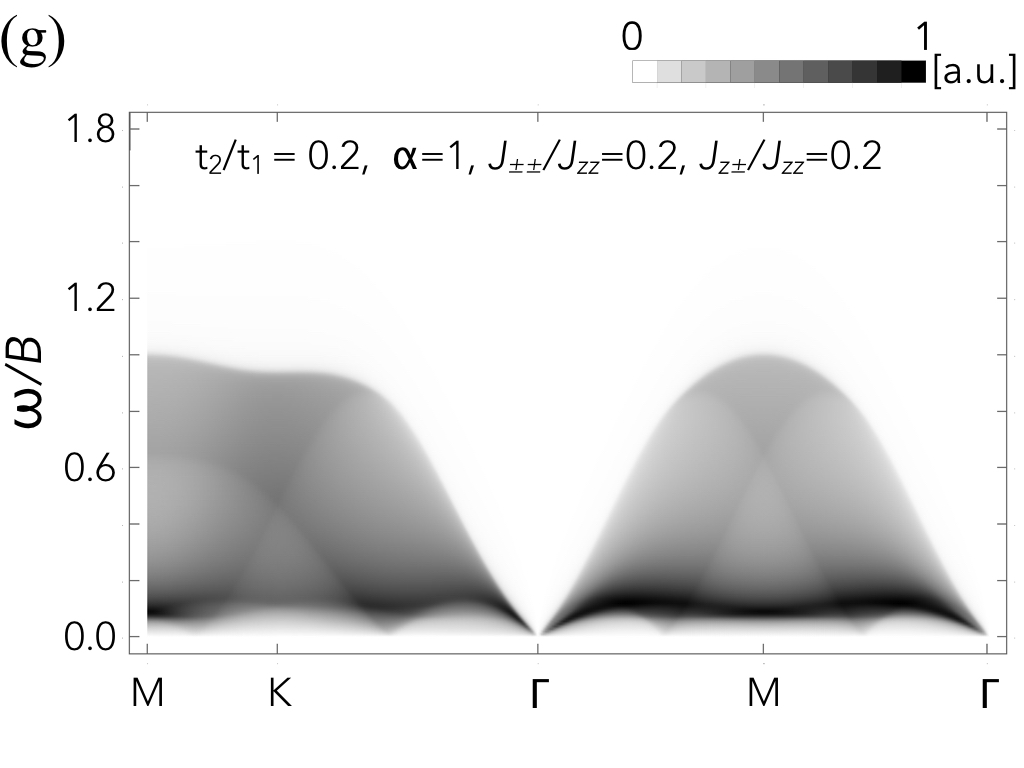}
\includegraphics[width=.24\textwidth]{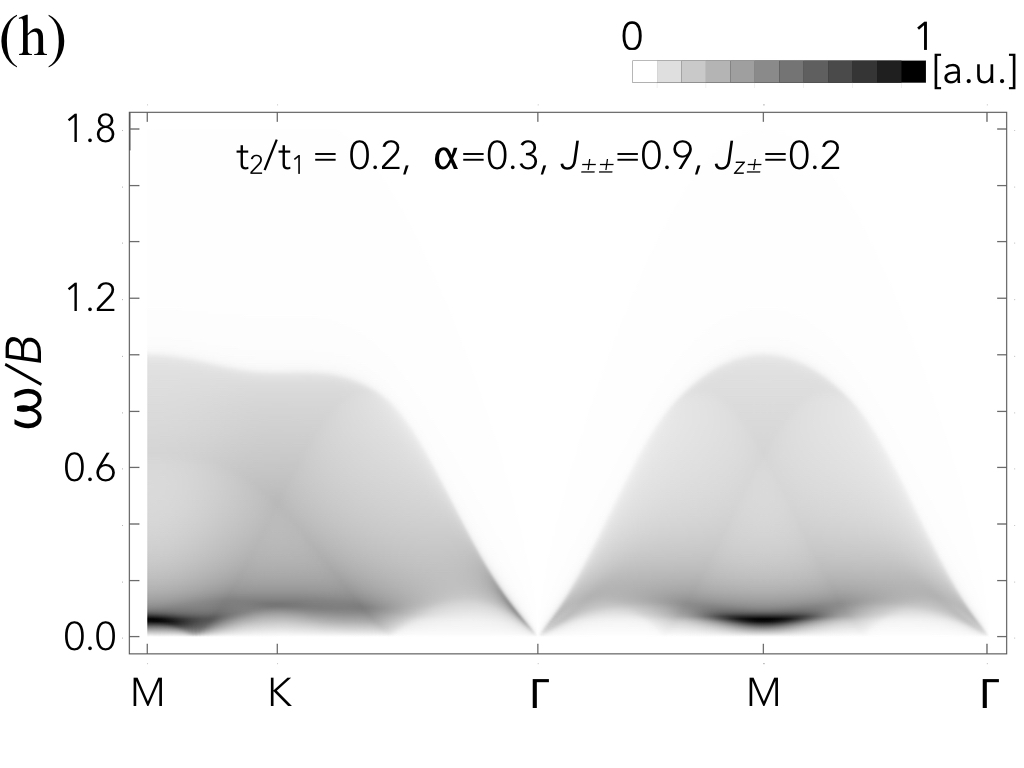}
\caption{(a-e) Dynamic spin structure factor for the free spinon theory 
of the U1A00 state with different values of $t_2/t_1$. (f-h) 
The evolution of $S^{\rm RPA}({\boldsymbol q}, \omega)$ as a 
function of $J_{\pm\pm}$. In all subfigures, the energy transfer 
is normalized against the corresponding bandwidth $B$. The parameter 
$\alpha$ is defined as $J_{zz}/t_1$. }
\label{sfig2}
\end{figure*}

The spinon mean-field Hamiltonian of the U1A00 state is
\begin{equation}
H_{\text{MF}}^{\text{U1A00}} = -t_1 
\sum_{\langle \boldsymbol{r} \boldsymbol{r}' \rangle, \alpha }
f^{\dagger}_{\boldsymbol{r}\alpha} f^{}_{\boldsymbol{r}\alpha}
-t_2 
\sum_{\langle\langle \boldsymbol{r} \boldsymbol{r}' \rangle\rangle, \alpha}
f^{\dagger}_{\boldsymbol{r}\alpha} f^{}_{\boldsymbol{r}\alpha} ,
\label{t1t2eq}
\end{equation}
from which we compute the dynamic spin structure factor for different
choices $t_2/t_1$. The dynamic spin structure factor is given by
\begin{eqnarray}
{\mathcal S}({\boldsymbol q}, \omega) &=& 
\frac{1}{N} \sum_{ {\boldsymbol r} , {\boldsymbol r}' }
e^{i {\boldsymbol q} \cdot ({\boldsymbol r} - {\boldsymbol r}')} 
\int  dt\, e^{-i \omega t }
\nonumber \\
&& \,
{ \langle \Psi_{\text{MF}}^{\text{U1A00}} |}
S^-_{\boldsymbol r} (t) S^+_{{\boldsymbol r}'} (0) 
{| \Psi_{\text{MF}}^{\text{U1A00}} \rangle}
\nonumber \\ 
&=& \sum_{n} \delta ( \omega - \xi_{n{\boldsymbol q}} ) \,
|\langle n | S^+_{\boldsymbol q } |\Psi_{\text{MF}}^{\text{U1A00}} \rangle|^2,
\end{eqnarray}
where $N$ is the total number of spins, the summation is over
all mean-field states with the spinon particle-hole excitation, 
$\xi_{n{\boldsymbol q}}$ is the energy of the $n$-th excited 
state with the momentum ${\boldsymbol q}$. 
The results are depicted in Fig.~\ref{sfig2}(a-e)
and are consistent with the inelastic neutron scattering 
results~\cite{YaoShenNature,Martin2016}. All the results 
so far are {\it independent} from any microscopic spin 
interaction. 
 
\subsection{Variational calculation and random phase approximation}

Here we consider the microscopic spin Hamiltonian that 
was introduced in Refs.~\onlinecite{YueshengPRL,YaodongPRB},
\begin{eqnarray}
H_{\text{spin}} &=& \sum_{\langle {\boldsymbol r} {\boldsymbol r}' \rangle}
J_{zz}^{} S^z_{\boldsymbol r} S^z_{{\boldsymbol r}'}
+{ J_{\pm}^{} ( S^+_{\boldsymbol r} S^-_{{\boldsymbol r}'}
+  S^-_{\boldsymbol r} S^+_{{\boldsymbol r}'})} \nonumber \\
&&
+ {J_{\pm\pm}^{} (\gamma_{{\boldsymbol r} {\boldsymbol r}' }^{}
S^+_{\boldsymbol{r}} S^+_{\boldsymbol{r}'} +
\gamma_{{\boldsymbol r} {\boldsymbol r}' }^{\ast}
S^-_{\boldsymbol{r}} S^-_{\boldsymbol{r}'} )} \nonumber \\
&&
{- \frac{i}{2} J_{z\pm}^{} [  (\gamma^{\ast}_{{\boldsymbol r}{\boldsymbol r}'}
S^+_{\boldsymbol r} - \gamma^{}_{{\boldsymbol r}{\boldsymbol r}'} S^-_{\boldsymbol r})}
{S_{\boldsymbol{r}'}^z }
\nonumber \\
&&
+ S^z_{\boldsymbol{r}} (\gamma^{\ast}_{{\boldsymbol r}{\boldsymbol r}' }
S^+_{{\boldsymbol r}'} -\gamma^{}_{{\boldsymbol r}{\boldsymbol r}' }
S^-_{{\boldsymbol r}'} ) ],
\end{eqnarray}
where $\gamma_{{\boldsymbol r}{\boldsymbol r}'} = 1, e^{i 2\pi/3}, e^{-i 2\pi/3}$ 
for ${{\boldsymbol r}{\boldsymbol r}'}$ along the ${\boldsymbol a}_1, {\boldsymbol a}_2$
and ${\boldsymbol a}_3$ bonds, respectively. Here, ${\boldsymbol a}_3 = - {\boldsymbol a}_1
-{\boldsymbol a}_2$. 
It was suggested and demonstrated that the anisotropic 
$J_{\pm\pm}$ and $J_{z\pm}$ interactions compete with the 
XXZ part of the Hamiltonian and may lead to 
disordered state~\cite{YueshengPRL,YaodongPRB,Yaodong201608}.
Our calculation does show the enhancement of 
quantum fluctuation in certain regions of the phase 
diagram~\cite{YaodongPRB}. Here we comment about the choices 
of the exchange couplings in the main text and 
in the following calculation. The $J_{zz}$ and $J_{\pm}$ couplings 
can be determined by the Curie-Weiss temperature measurement 
on a single crystal sample. The complication comes from the 
subtraction of the Van Vleck susceptibility. Due to the Ga$^{3+}$/Mg$^{2+}$ 
exchange disorder in the non-magnetic layers, although these ions do not directly
enter the Yb exchange path, it may modify the local crystal
field environment of the Yb$^{3+}$ ion and thus lead to some complication
and variation of the Van Vleck susceptibility. As a result, the {\it very precise}
determination of the $J_{zz}$ and $J_{\pm}$ couplings can be an issue. 
That may explain some differences of the $J_{zz}$ and $J_{\pm}$ couplings 
that were obtained~\cite{YueshengPRL,YaodongPRB,YaoShenNature,Martin2016,Yaodong201608}.
Partly for the same reason, the results on $J_{\pm\pm}$ and $J_{z\pm}$ may 
also be affected. However, quantum spin liquid, if it exists as the ground state of 
our generic model, is expected to be a phase that covers a finite region of 
the phase diagram. Therefore, the {\it very precise} value of the couplings
may not be quite necessary from this point of view. Therefore, we here rely
on our previous results of the quantum fluctuation for the mean-field phase 
diagram that indicates strong fluctations in certain parameter regimes.
We choose the exchange parameters from these disordered regions. 

For this spin Hamiltonian, the mean-field variational energy is 
given as
\begin{widetext}
\begin{eqnarray}
    E_{\rm var} & = &  
    \bra{\Psi_{\text{MF}}^{\text{U1A00}}} 
    H_{\rm spin}^{} 
    \ket{\Psi_{\text{MF}}^{\text{U1A00}} } 
    = \frac{1}{L^2} \sum_{\boldsymbol q} 
    \bra{\Psi_{\text{MF}}^{\text{U1A00}}} 
    J_{zz}({\boldsymbol q}) S^z_{\boldsymbol q} S^z_{-{\boldsymbol q}} 
    + 2 J_{\pm}({\boldsymbol q}) S^+_{\boldsymbol q} S^-_{-{\boldsymbol q}} 
    \ket{\Psi_{\text{MF}}^{\text{U1A00}}  }  
    \nn
    &=& \frac{1}{L^2} \sum_{\boldsymbol q} \lz J_{zz}({\boldsymbol q}) 
    \sum_n \left| \bra{n} S^z_{\boldsymbol q} \ket{\Psi_{\text{MF}}^{\text{U1A00}} }  \right|^2 
    + 2 J_{\pm}({\boldsymbol q}) \sum_n \left| \bra{n} S^+_{\boldsymbol q} 
    \ket{\Psi_{\text{MF}}^{\text{U1A00}} }  \right|^2 \rz \nn
    &=&  \frac{1}{L^4}  \sum_{\boldsymbol q} \lz 
    \frac{J_{zz}({\boldsymbol q})}{4} \sum_{n, {\boldsymbol k}} \left| \bra{n} 
    f^\dg_{\boldsymbol{k+q}, \ua} f^\pg_{\boldsymbol k, \ua} 
    - f^\dg_{\boldsymbol{k+q}, \da} f^\pg_{\boldsymbol k, \da} 
    \ket{\Psi_{\text{MF}}^{\text{U1A00}} }  \right|^2
    + 2 J_{\pm}({\boldsymbol q})
     \sum_{n, {\boldsymbol k}} \left| \bra{n}  f^\dg_{\boldsymbol{k+q}, \ua} 
     f^\pg_{\boldsymbol k, \da} \ket{\Psi_{\text{MF}}^{\text{U1A00}} }  
     \right|^2 \rz ,
\end{eqnarray}
\end{widetext}
where we have omitted $J_{\pm\pm}$ and $J_{z\pm}$ because 
they do not conserve spin, therefore their contribution 
to $E_{\rm var}$ is zero. This is an artifact of the 
free spinon theory of $H_{\text{MF}}^{\text{U1A00}}$ that
only includes isotropic spinon hoppings for the first 
two neighbors.

Due to the isotropic spinon hoppings, $H_{\text{MF}}^{\text{U1A00}}$
does not explicitly reflect the absence of spin-rotational symmetry 
that is brought by the $J_{\pm\pm}$ and $J_{z\pm}$ interactions. 
To incorporate the $J_{\pm\pm}$ and $J_{z\pm}$ interactions, as we describe
in the main text, we followed the phenomenological treatment for the ``$t$-$J$'' 
model in the context of cuprate superconductors~\cite{PhysRevLett.82.2915}
and consider $H = H_{\text{MF}}^{\text{U1A00}} + H_{\text{spin}}'$,
where $H_{\text{spin}}'$ are the 
$J_{\pm\pm}$ and $J_{z\pm}$ interactions. 
In the parton construction, $H_{\text{spin}}'$
is treated as the spinon interactions and thus
introduces the spin rotational symmetry breaking. 
With a random phase approximation for 
the interaction $H_{\text{spin}}'$,  
we obtain the dynamic spin 
susceptibility~\cite{PhysRevLett.82.2915}
\begin{eqnarray}
\chi^{\rm RPA}({\boldsymbol q}, \omega) = \left[ {\bf 1} 
- \chi^0 ({\boldsymbol q}, \omega) {\mathcal J}({\boldsymbol{q}}) 
\right]^{-1} \chi^0({\boldsymbol q}, \omega), 
\label{chiRPA}
\end{eqnarray}
where $\chi^0$ is the free-spinon susceptibility, and ${\cal J}({\bf q})$ is 
the spin exchange matrix from $H^\p_{\rm spin}$, 
\begin{eqnarray}
    &&{\cal J}({\boldsymbol q}) = \nonumber 
    \\
    &&\begin{pmatrix}
     2(u_{\boldsymbol q} - v_{\boldsymbol q}) J_{\pm\pm}
     & -2 \sqrt{3} w_{\boldsymbol q} J_{\pm\pm}
     & -\sqrt{3} w_{\boldsymbol q} J_{z\pm}  \\
     -2 \sqrt{3} w_{\boldsymbol q} J_{\pm\pm} &
     2 (-u_{\boldsymbol q} + v_{\boldsymbol q}) J_{\pm\pm}
     & \left(u_{\boldsymbol q} - v_{\boldsymbol q}  \right) J_{z\pm} \\
     -\sqrt{3} w_{\boldsymbol q} J_{z\pm}
     & \left(u_{\boldsymbol q} - v_{\boldsymbol q} \right) J_{z\pm}
     &  
     0
    \end{pmatrix} \quad \quad
\end{eqnarray}
with $u_{\boldsymbol q} = \cos({\boldsymbol q}\cdot {\boldsymbol a}_1)$, $v_{\boldsymbol q} = \frac{1}{2} \left( \cos({\boldsymbol q}\cdot {\boldsymbol a}_2) + \cos({\boldsymbol q}\cdot {\boldsymbol a}_3) \right)$, and $w_{\boldsymbol q} = \frac{1}{2} \left( \cos({\boldsymbol q}\cdot {\boldsymbol a}_2) - \cos({\boldsymbol q}\cdot {\boldsymbol a}_3) \right)$.
The renormalized ${\mathcal S}^{\text{RPA}}({\boldsymbol q}, \omega)$
can be read off from $\chi^{\text{RPA}}$ via 
${\cal S}^{\rm RPA}({\boldsymbol q}, \omega) = 
-\frac{1}{\pi}{\rm Im} \left[ \chi^{\rm RPA}
({\boldsymbol q}, \omega) \right]^{+-}$ and is plotted in Fig.~\ref{fig3}(b)
in the main text. 

The very precise values of $J_{\pm\pm}$ and $J_{z\pm}$ 
cannot be determined from the existing {\it data-rich} 
neutron scattering experiment in a strong field normal 
to the triangular plane. This is partly due to the 
experimental resolution, and is also due to the fact that  
the linear spin wave spectrum for the field normal to the plane 
is {\it independent} of $J_{z\pm}$ and is not quite 
sensitive to $J_{\pm\pm}$~\cite{YaodongPRB,Yaodong201608}. 
In Fig.~\ref{fig3}(b) of the main text, instead, we choose ${(J_{\pm\pm},J_{z\pm})}$
to fall into the disordered region of the phase diagram 
in Ref.~\onlinecite{YaodongPRB} where the quantum fluctuations 
are expected to be strong~\cite{YaodongPRB}.

\section{The U1B states}

\begin{figure*}[thp]
\centering
\includegraphics[width=.32\textwidth]{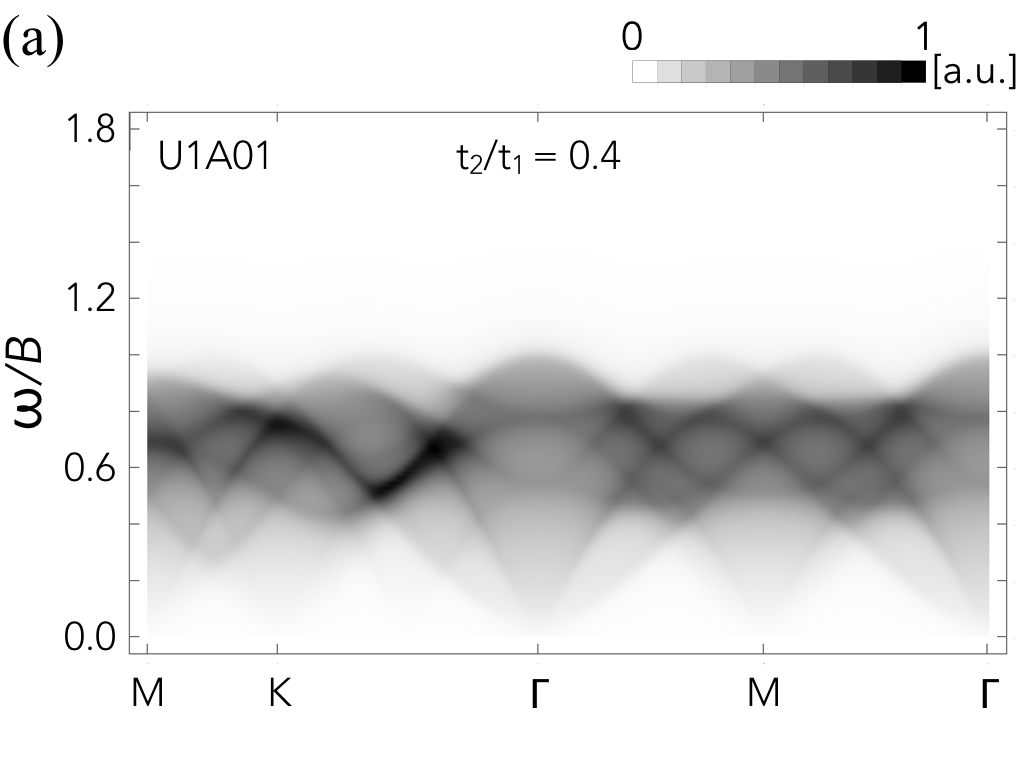}
\includegraphics[width=.32\textwidth]{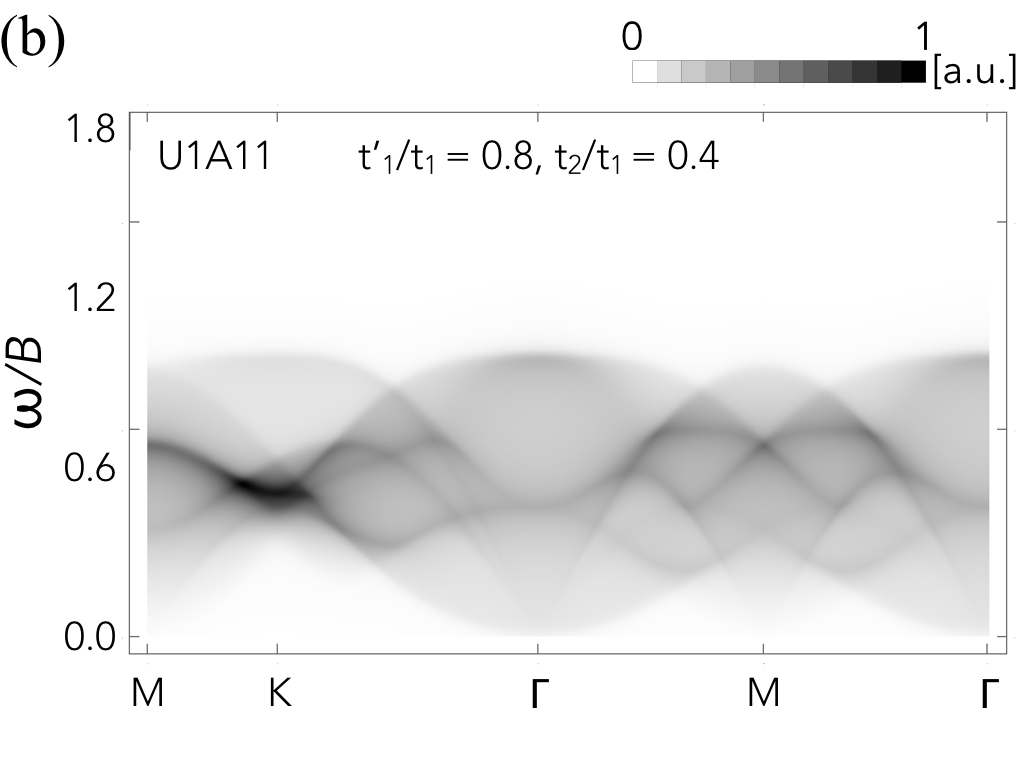}
\includegraphics[width=.32\textwidth]{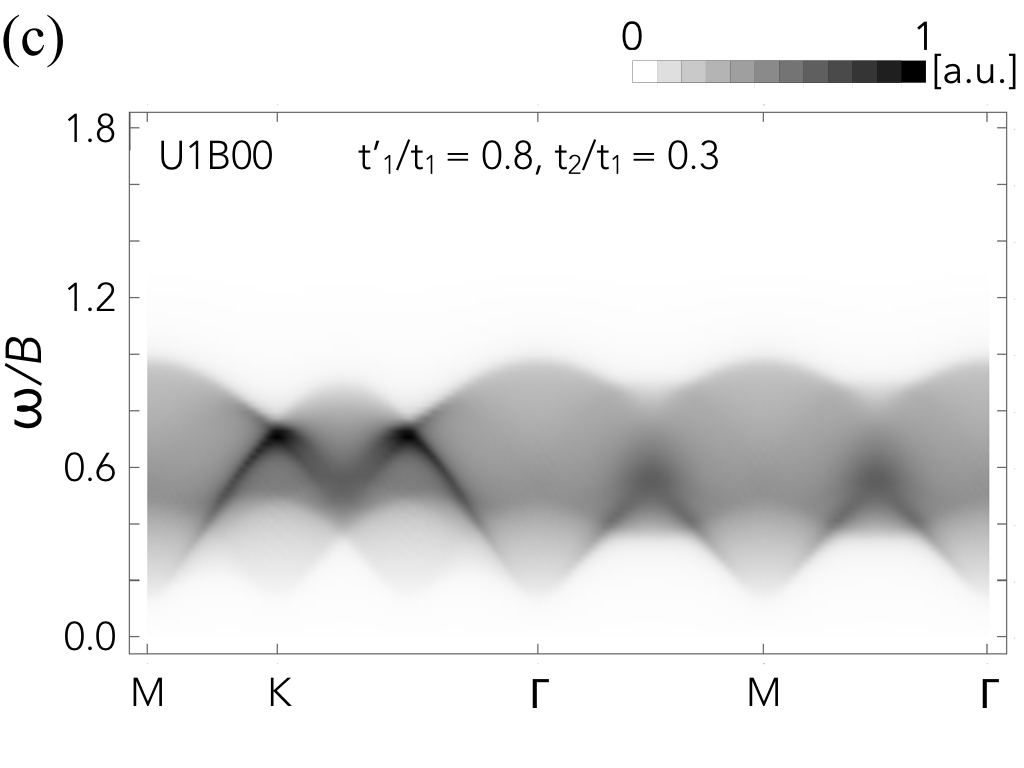}
\includegraphics[width=.32\textwidth]{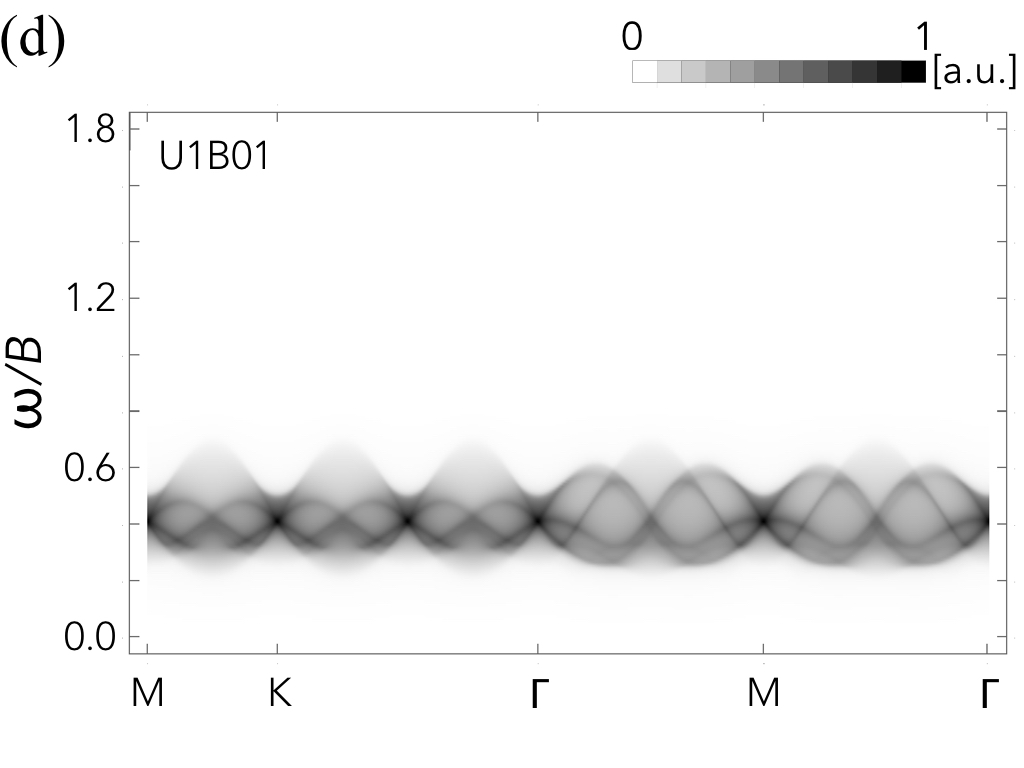}
\includegraphics[width=.32\textwidth]{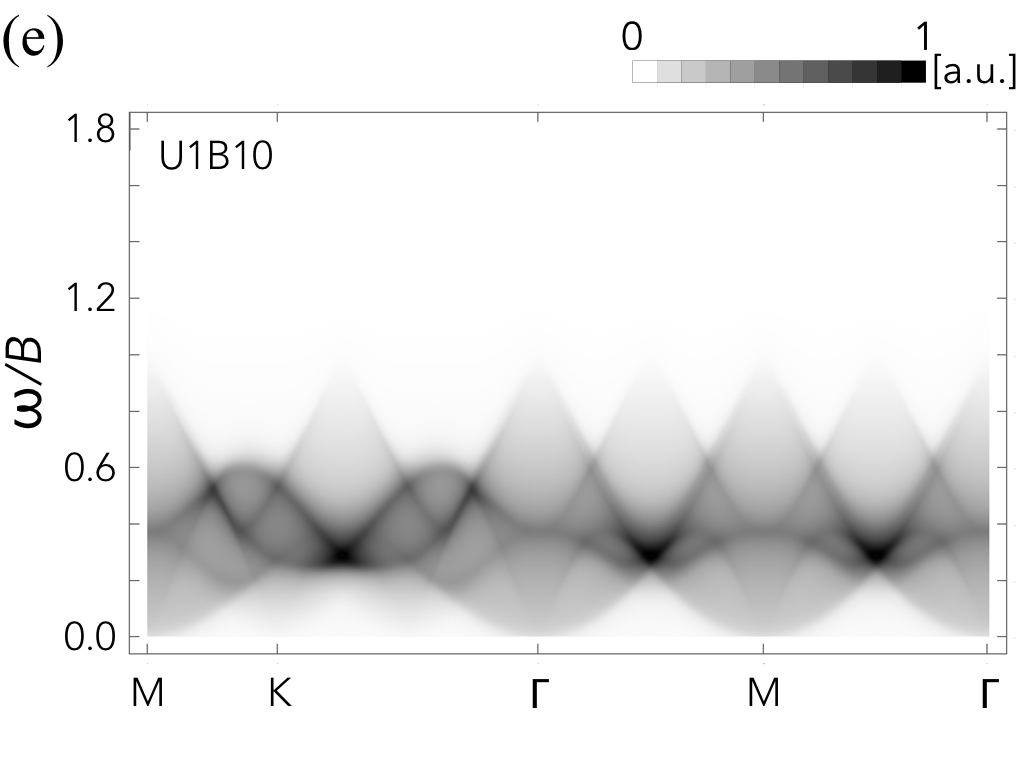}
\includegraphics[width=.32\textwidth]{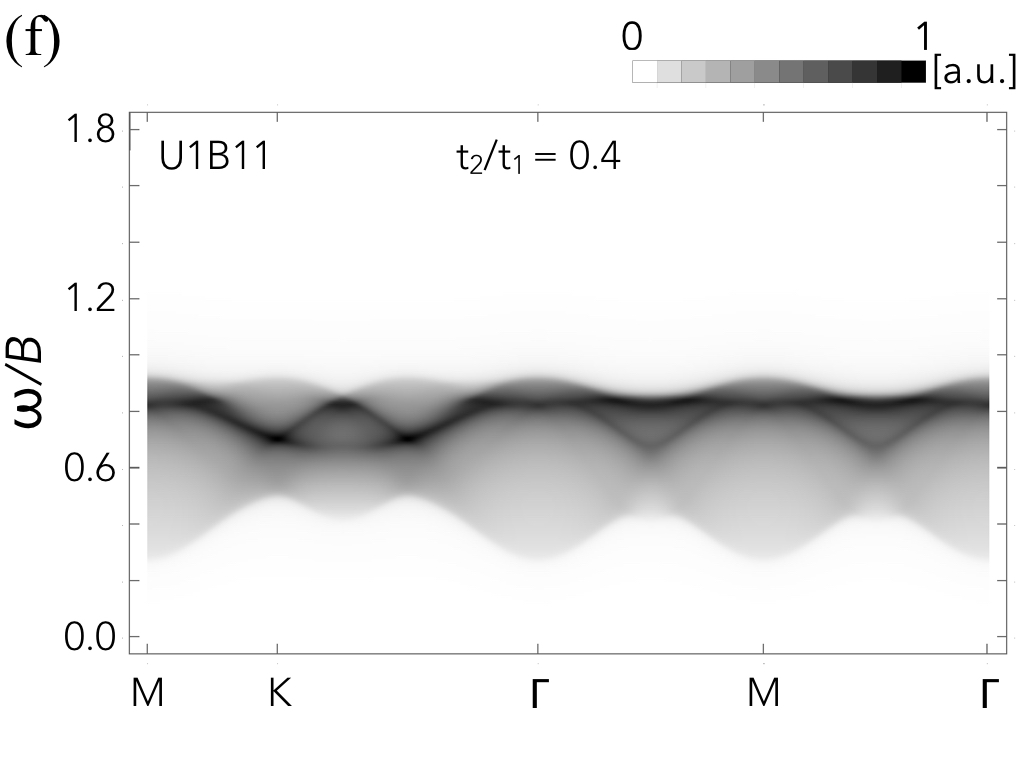}
\caption{Dynamic spin structure factor for six free spinon mean-field states 
other than U1A00. Note the U1A10 Hamiltonian is identically zero for the first 
and second neighbor hoppings. None of them is consistent with the spinon 
Fermi surface picture. In all subfigures, the energy transfer is normalized 
against the corresponding bandwidth $B$.}
\label{sfig3}
\end{figure*}

In this section we use PSG to determine the free spinon mean-field Hamiltonian 
for the U1B states to the first and second spinon hoppings. 
In Fig.~\ref{sfig3}, we further present their spectroscopic 
features for comparison. 
Like the notation for U1As, the U1B states are also labeled 
by U1B$n_{C_2}n_{S_6}$.

\subsection{The U1B00 state}

For the $\pi$-flux states, the dynamic spin structure factor has an
enhanced periodicity due to anticommutative lattice translations. 
One direct consequence of the periodicity is that $\Gamma$ and M 
become equivalent, and the V-shaped upper excitation edge in 
Ref.~\onlinecite{YaoShenNature} cannot be reproduced for the 
U1B states.

We choose the spinon basis in the momentum space 
$f_{{\boldsymbol k}, I} = (f^\pg_{A, {\boldsymbol k}, \ua}, 
f^\pg_{B, {\boldsymbol k}, \ua}, f^\pg_{A, {\boldsymbol k}, \da}, 
f^\pg_{B, {\boldsymbol k}, \da})^T$, 
where $A$ and $B$ denote the two inequivalent sites 
in each unit cell due to the $\pi$ flux.

The Hamiltonian is written in terms of the Dirac matrices 
$\Gamma^a$ and their anticommutators 
\begin{equation}
\Gamma^{ab} = [\Gamma^a, \Gamma^b]/(2i). 
\end{equation}
The representation is chosen to be $\Gamma^{(1,2,3,4,5)} = (\sigma^x\otimes{\bf 1}, \sigma^z\otimes {\bf 1}, \sigma^y\otimes\tau^x, \sigma^y\otimes\tau^y, \sigma^y\otimes\tau^z)$. 
$\Gamma^a$ and $\Gamma^{ab}$ is odd under time reversal except 
when $a=4$ or $b=4$. The Hamiltonian is thus
\begin{eqnarray}
    h({\bf k}) = \sum_{a=1}^5 d_a({\boldsymbol k})\Gamma^a 
    + \sum_{a<b=1}^5 d_{ab}({\boldsymbol k}) \Gamma^{ab}. 
\end{eqnarray}

For the U1B00 state, we have 
\begin{eqnarray}
    d_3({\boldsymbol k}) &=& t_1^\p \sin ( {k_x}/{2}-{\sqrt{3} k_y}/{2} ) ,\nn
    d_4({\boldsymbol k}) &=& t_1^\p \cos ( {k_x}/{2}+{\sqrt{3} k_y}/{2} ) ,\nn
    d_5({\boldsymbol k}) &=& -2 t_1^\p \sin (k_x) ,\nn
    d_{13}({\boldsymbol k}) &=& -2 t_1 \sin ( {k_x}/{2}-{\sqrt{3} k_y}/{2} ), \nn
    d_{14}({\boldsymbol k}) &=& -2 t_1 \cos ( {k_x}/{2}+{\sqrt{3} k_y}/{2} ), \nn
    d_{15}({\boldsymbol k}) &=& -2 t_1 \sin (k_x), \nn
    d_{23}({\boldsymbol k}) &=& -\sqrt{3} t_1^\p \sin ( {k_x}/{2}- {\sqrt{3} k_y}/{2} ),  \nn
    d_{24}({\boldsymbol k}) &=& \sqrt{3} t_1^\p \cos ( {k_x}/{2}+ {\sqrt{3} k_y}/{2} ) , \nn
    d_{34}({\boldsymbol k}) &=& 2 t_2 \cos ( \sqrt{3} k_y ) ,\nn
    d_{35}({\boldsymbol k}) &=& 2 t_2 \sin ( {3 k_x}/{2} - {\sqrt{3} k_y}/{2} ) , \nn
    d_{45}({\boldsymbol k}) &=& 2 t_2 \cos ( {3 k_x}/{2} + {\sqrt{3} k_y}/{2} ) .
\end{eqnarray}

\subsection{The U1B01 state}
\begin{eqnarray}
d_3({\boldsymbol k}) &=&   t_2 \sin ( {3 k_x}/{2} + {\sqrt{3} k_y}/{2} ) ,  \nn
d_4({\boldsymbol k}) &=& - t_2 \cos ( {3 k_x}/{2}- {\sqrt{3} k_y}/{2} )  ,\nn
d_5({\boldsymbol k}) &=& 2 t_2 \sin ( \sqrt{3} k_y ) , \nn
d_{23}({\boldsymbol k}) &=& -\sqrt{3} t_2 \sin ( {3 k_x}/{2}+ {\sqrt{3} k_y}/{2} ), \nn
d_{24}({\boldsymbol k}) &=& -\sqrt{3} t_2 \cos ( {3 k_x}/{2}- {\sqrt{3} k_y}/{2} ).
\end{eqnarray}
 
\subsection{The U1B10 state}
\begin{eqnarray}
d_3({\boldsymbol k}) &=& - \sqrt{3} t_1 \sin \big[    ( k_x - \sqrt{3} k_y ) /2 \big], \nn
d_4({\boldsymbol k}) &=& \sqrt{3} t_1 \cos \big[   ( k_x+\sqrt{3} k_y )  /2 \big], \nn
d_{23}({\boldsymbol k}) &=& - t_1 \sin \big[ ( k_x-\sqrt{3} k_y ) /2 \big], \nn
d_{24}({\boldsymbol k}) &=& - t_1 \cos \big[ ( k_x+\sqrt{3} k_y ) /2 \big] , \nn
d_{25}({\boldsymbol k}) &=& 2 t_1 \sin k_x.
\end{eqnarray}

\subsection{The U1B11 state}

\begin{eqnarray}
    d_3({\boldsymbol k}) &=& -\sqrt{3} t_2 \sin \big[ ( 3  k_x+\sqrt{3}
   k_y) /2   \big] ,\nn
    d_4({\boldsymbol k}) &=& -\sqrt{3} t_2 \cos \big[   ( 3 k_x-\sqrt{3}
   k_y ) /2  \big], \nn
    d_{23}({\boldsymbol k}) &=& -t_2 \sin \big[ ( 3 k_x+\sqrt{3} k_y ) /2 \big] ,\nn
    d_{24}({\boldsymbol k}) &=& t_2 \cos \big[  ( 3 k_x-\sqrt{3} k_y ) /2 \big] ,\nn
    d_{25}({\boldsymbol k}) &=& -2 t_2 \sin ( \sqrt{3} k_y ) ,\nn
    d_{34}({\boldsymbol k}) &=& 2 t_1 \cos (k_x) , \nn
    d_{35}({\boldsymbol k}) &=& -2 t_1 \sin  \big[  ( k_x+\sqrt{3} k_y ) /2 \big], \nn
    d_{45}({\boldsymbol k}) &=& -2 t_1 \cos  \big[ ( k_x-\sqrt{3} k_y ) /2 \big] .
\end{eqnarray}

\bibliography{Refs_Nov}

\end{document}